\documentstyle[titlepage,12pt]{article}
\newcommand{\beq}{\begin{equation}}
\newcommand{\eeq}{\end{equation}}
\newcommand{\eq}[1]{eq.(\ref{#1})}
\title {
\begin{flushright} PSU/TH/142
\end{flushright}
An $\alpha^{2}(Z \alpha)^{5}m$ Contribution to the Hydrogen Lamb Shift
\\
from Virtual Light by Light Scattering}
\author
{Michael I. Eides \thanks{E-mail address: eides@lnpi.spb.su}\\
Petersburg Nuclear Physics Institute,\\
\medskip
Gatchina, St.Petersburg 188350, Russia,\\
\and
Howard Grotch\thanks{E-mail address: h1g@psuvm.psu.edu}\\
Department of Physics, Pennsylvania State University,\\
\medskip University Park, Pennsylvania 16802, USA\\
and
\and
Peter Pebler \thanks{E-mail address: pebler@minerva.cis.yale.edu}\\
Department of Physics, Yale University,\\
New Haven, CT 06520, USA}
\date{February 15, 1994}

\begin{document}
\baselineskip-15cm
\maketitle
\lineskip3pt
\begin{abstract}
The radiative correction to the Lamb shift of order
$\alpha^{2}(Z\alpha)^5m$ induced by the light by light scattering insertion
in external photons is obtained. The new contribution turns out to be equal to
$-0.122(2)\alpha^2(Z\alpha)^5/(\pi n^3)(m_r/m)^3m$.  Combining this
contribution with our previous results we obtain the complete correction of
order $\alpha^{2}(Z\alpha)^5m$ induced by all diagrams with  closed electron
loops. This correction is $37.3(1)$ kHz and $4.67(1)$ kHz for
the $1S$- and $2S$-states in hydrogen, respectively.

\end{abstract}

\section{INTRODUCTION}

Recently we started a calculation of all contributions to the Lamb shift
of order $\alpha^{2}(Z\alpha)^5m$. It was  shown that there  exist  six
gauge  invariant sets of diagrams, which produce such corrections. All these
diagrams may be obtained by different dressings from the skeleton diagram
which contains two external photons attached to the electron line.
Contributions induced by polarization operator insertions in external
photons, by simultaneous insertion of a radiative photon in electron line and
a one-loop polarization operator in the external photons and by polarization
operator insertions in radiative photons have been calculated in
\cite{ego,eg3,eg4}.

We present below a calculation of the $\alpha^{2}(Z\alpha)^5m$ contribution to
the Lamb shift in  hydrogenlike  ions, induced by penultimate gauge
invariant set of graphs.  These are the graphs  containing  one-loop  light
by light scattering insertions between two external photon lines (see Fig.$1$).

The contribution to the Lamb shift, produced by the diagrams in Fig.$1$ is
given
by  the  expression  (see, e.g. \cite{ego})

\beq   \label{lamb}
\Delta E=\frac{\alpha^2(Z\alpha)^5}{\pi n^3}m
(\frac{m_r}{m})^3\frac{1}{\pi^2}\int\frac{d^4q}{\pi^2i}\frac{1}{q^4}
\int\frac{d^3{\bf k}}{4\pi{\bf k}^4}({\cal A}_{\alpha\beta}
+{\cal B}_{\alpha\beta}){\cal S}_{\alpha\beta 00},
\eeq

where $q_\mu$ is the four-momentum  of  the  upper  photon
lines, $k_{\mu}=(0,{\bf k})$ is the spatial momentum  of  external
photons, ${\cal A}_{\alpha\beta}$ and ${\cal B}_{\alpha\beta}$ are
electron-line factors, and ${\cal S}_{\alpha\beta 00}$ is the light by light
scattering tensor\footnote {Really ${\cal S}_{\alpha\beta 00}$ is
defined as one fourth of the light by light scattering tensor; see
normalization conventions for this tensor as well as for the electron factor
in the next section.}.

\section{SIMPLIFICATION OF THE EXPRESSION FOR THE ENERGY SHIFT}

We begin the calculation with a simplification of the integrand in  \eq{lamb}.
The general expression for the light by light scattering tensor is

\beq       \label{lghtbylght}
{\cal S}_{\alpha\beta 00}=\int\frac{d^4p}{\pi^2 i}(2L_{\alpha\beta
00} +C_{\alpha\beta 00})\equiv \int\frac{d^4p}{\pi^2
i}S_{\alpha\beta 00},
\eeq

where

\beq                  \label{ladlght}
L_{\alpha\beta 00}=-\frac{1}{4D_1^2D_2D_3}Tr\{\gamma_0(\hat p-\hat k+
m)\gamma_0(\hat p+m)\gamma_\beta(\hat p-\hat q+m)\gamma_\alpha(\hat p+m)\},
\eeq

corresponds to each of the ladder diagrams {\it a} and {\it b}
in Fig.$1$, and

\beq               \label{croslght}
C_{\alpha\beta 00}=
\eeq
\[
-\frac{1}{4D_1D_2D_3D_4}Tr\{\gamma_0(\hat p-\hat k+
m)\gamma_\beta(\hat p-\hat k-\hat q+m)\gamma_0(\hat p-\hat
q+m)\gamma_\alpha(\hat p+m)\}
\]

corresponds to the crossed diagram {\it c} in Fig.$1$. Factors in the
denominators above are defined as follows

\beq
D_1=p^2-m^2,\: D_2=(p-k)^2-m^2, \: D_3=(p-q)^2-m^2, \:
D_4=(p-k-q)^2-m^2.
\eeq

The explicit expression for the electron factor in \eq{lamb} is the  sum of two
terms, where

\beq            \label{elpar}
{\cal A}_{\alpha\beta}=\frac{Tr\{(\gamma_0+1)\gamma_\alpha(\gamma_0 m+\hat
q+m)\gamma_\beta\}}{4(q^2+2mq_0)}m^2
\eeq
\[
=  \frac{-g_{\alpha\beta}q_0 + 2g_{\alpha 0}g_{\beta 0}m + g_{\alpha
0}q_\beta + q_\alpha g_{\beta 0}}{q^2+2mq_0}m^2\equiv\frac{A_{\alpha\beta}}
{q^2+2mq_0}
\]
corresponds to the diagram with nonintersecting  upper photon lines and

\beq
{\cal B}_{\alpha\beta}=\frac{Tr\{(\gamma_0+1)\gamma_\beta(\gamma_0 m-\hat
q+m)\gamma_\alpha\}}{4(q^2-2mq_0)}m^2
\eeq
\[
=\frac{g_{\alpha\beta}q_0 + 2g_{\alpha 0}g_{\beta 0}m -
g_{\alpha 0}q_\beta - q_\alpha g_{\beta
0}}{q^2-2mq_0}m^2\equiv\frac{B_{\alpha\beta}} {q^2-2mq_0}
\]
corresponds to the diagram with crossed upper photon lines \footnote{We
have included an additional factor of $m^2$ in the definition of the electron
 line factor simply to preserve the appearance of the expression for the energy
shift in \eq{lamb} even after transition to the dimensionless
momenta.}.

Note that  ${\cal A}_{\alpha\beta}(q)={\cal B}_{\alpha\beta}(-q)$ and both
${\cal A}_{\alpha\beta}$  and ${\cal B}_{\alpha\beta}$ are symmetric
relative to permutation of indices. Hence, only the even in $q$ and symmetric
under permutation of indices $\alpha$ and $\beta$ part of the light by light
scattering tensor is relevant for calculations below. Due to gauge
invariance terms in the light by light scattering tensor which are
proportional to vectors $q_\alpha$ and $q_\beta$ are also irrelevant for
calculation of the contribution to the Lamb shift and we omit such terms
below.

Since ${\cal A}_{\alpha\beta}(q)={\cal B}_{\alpha\beta}(-q)$ the expression
for the energy shift may be presented in the following form

\beq   \label{lamb2}
\Delta E=\frac{\alpha^2(Z\alpha)^5}{\pi n^3}m
(\frac{m_r}{m})^3\frac{1}{\pi^2}\int\frac{d^4q}{\pi^2i}
\frac{1}{q^4}\int\frac{d^3{\bf k}}{4\pi{\bf k}^4}{\cal A}_{\alpha\beta}(q)
[{\tilde{\cal S}}_{\alpha\beta 00}(q)+{\tilde{\cal S}}_{\alpha\beta
00}(-q)]
\eeq
\[
=\frac{\alpha^2(Z\alpha)^5}{\pi n^3}m
(\frac{m_r}{m})^3\frac{1}{\pi^2}\int\frac{d^4q}{\pi^2i}
\frac{1}{q^4}\int\frac{d^3{\bf k}}{4\pi{\bf k}^4} \times
\]
\[
\int\frac{d^4p}{\pi^2 i} {\cal
A}_{\alpha\beta}(q)[{\tilde S}_{\alpha\beta 00}(q,k,p)+{\tilde
S}_{\alpha\beta 00}(-q,k,p)],
\]
where ${\tilde{\cal S}}_{\alpha\beta 00}$ is the integrand of the light by
light scattering tensor symmetrized in $\alpha$ and $\beta$.

Both $k$ and $p$ are dummy integration momenta and we use
substitutions $k\rightarrow -k$ and  $p\rightarrow -p$ while calculating
the second product in the integrand in \eq{lamb2}. After these substitutions
denominators in the integral representation for the light by light
scattering tensor with external momenta of opposite sign return to the
original form (see \eq{ladlght} and \eq{croslght}). Moreover, it is easy to
see that

\beq
{\tilde S}_{\alpha\beta 00}(-q,-k,-p;m) ={\tilde S}_{\alpha\beta
00}(q,k,p;-m),
\eeq
if one recollects that there exists an additional argument $m$ in the light
by light scattering tensor and takes into account explicit expressions
in \eq{ladlght} and \eq{croslght}. Traces in the numerators on
the right  hand sides in these equations convert these numerators
into even polynomials of $m$ and, hence, the expression
for the energy splitting in \eq{lamb2} reduces to

\beq   \label{lamb3}
\Delta E=
\eeq
\[
\frac{\alpha^2(Z\alpha)^5}{\pi n^3}m
(\frac{m_r}{m})^3\frac{2}{\pi^2}\int\frac{d^4q}{\pi^2i}
\frac{1}{q^4}\int\frac{d^3{\bf k}}{4\pi{\bf k}^4}
\int\frac{d^4p}{\pi^2 i} {\cal
A}_{\alpha\beta}(q) {\tilde S}_{\alpha\beta 00}(q,k,p;m).
\]

\section{CALCULATION OF THE LIGHT BY LIGHT SCATTERING  TENSOR}

\subsection{LADDER DIAGRAM}

The explicit expression for the tensor in \eq{ladlght} symmetrized over indices
 $\alpha$ and $\beta$, has the form

\beq
{\tilde L}_{\alpha\beta 00}=-\frac{1}{D_1^2D_2D_3}\{ - (kp)p^2
g_{\alpha\beta}
+ 2(kp)(pq)g_{\alpha\beta}+ 4(kp)p_{\alpha}p_{\beta}
\eeq
\[
+ (kp)g_{\alpha\beta}m^2 - (kq)p^2g_{\alpha\beta} + (kq)g_{\alpha\beta}m^2
- k_{\alpha}p^2p_{\beta} + k_{\alpha}p_{\beta}
m^2 - k_{\beta}p^2p_{\alpha} + k_{\beta}p_{\alpha}m^2
\]
\[
+ p^4g_{\alpha\beta} -
p^2(pq)g_{\alpha\beta} - 2p^2p_0^2g_{\alpha\beta} -
2p^2p_0p_{\alpha}g_{\beta 0} - 2p^2p_0p_{\beta}g_{\alpha 0} -
 2p^2p_0q_0g_{\alpha\beta} - 2p^2p_{\alpha}p_{\beta}
\]
\[
-2p^2g_{\alpha\beta}m^2 + 4(pq) p_0^2g_{\alpha\beta}+
(pq)g_{\alpha\beta}m^2 + 8p_0^2p_{\alpha}p_{\beta} + 2p_0^2
g_{\alpha\beta}m^2 +
2p_0p_{\alpha}g_{\beta 0}m^2
\]
\[
+ 2p_0p_{\beta}g_{\alpha 0}m^2 + 2p_0q_0g_{\alpha\beta}m^2
+ 2p_{\alpha}p_{\beta}m^2 + g_{\alpha\beta}m^4.
\]

Next we combine denominators with the help of the Feynman parameters $x$ and
$u$

\beq        \label{combp}
(1-u)D_1+u[xD_2+(1-x)D_3]=p'^2-\Delta,
\eeq
\[
p'=p-kux-qu(1-x),
\]
\[
\Delta=-k^2xu(1-ux)-q^2u(1-x)[1-u(1-x)]+2(kq)u^2x(1-x)+m^2,
\]
\[
\frac{1}{D_1^2D_2D_3}=6\int_0^1dx\int_0^1duu(1-u)\frac{1}{(p'^2-\Delta)^4}.
\]

Thus we obtain

\beq
{\tilde{\cal L}}_{\alpha\beta 00}=
\eeq
\[
-6\int_0^1dx\int_0^1duu(1-u)\int\frac{d^4p'}{\pi^2 i}\{
\frac{1}{3}p'^4{\tilde L}_{\alpha\beta 00}^{(0)}
+ \frac{1}{2}p'^2{\tilde L}_{\alpha\beta 00}^{(1)}
+ {\tilde L}_{\alpha\beta 00}^{(2)}\}
\frac{1}{(p'^2-\Delta)^4},
\]

where

\beq                \label{auxfunct}
{\tilde L}_{\alpha\beta 00}^{(0)}=- g_{\alpha 0}g_{\beta 0} +
g_{\alpha\beta},
\eeq

\medskip

\beq
{\tilde L}_{\alpha\beta 00}^{(1)}= - 2k^2g_{\alpha 0}g_{\beta 0}(ux)^2
+ 4k^2g_{\alpha\beta}(ux)^2 - k^2g_{\alpha\beta}(ux)
\eeq
\[
- 4(kq)g_{\alpha 0}g_{\beta 0}(ux)[u(1-x)]
+ 8(kq)g_{\alpha\beta}(ux)[u(1-x)] - (kq)g_{\alpha\beta}(ux)
\]
\[
- (kq)g_{\alpha\beta}[u(1-x)] - (kq)g_{\alpha\beta}
- 4k_\alpha k_\beta(ux)^2- 2q^2g_{\alpha 0}g_{\beta 0}[u(1-x)]^2
\]
\[
+ 4q^2g_{\alpha\beta}[u(1-x)]^2- q^2g_{\alpha\beta}[u(1-x)]
 - 4q_0^2g_{\alpha\beta}[u(1-x)]^2
\]
\[
- 2q_0^2g_{\alpha\beta} [u(1-x)] + 2g_{\alpha 0}g_{\beta 0}m^2 -
2g_{\alpha\beta}m^2,
\]

\medskip

\beq
{\tilde L}_{\alpha\beta 00}^{(2)}=k^4g_{\alpha\beta}(ux)^4 -
k^4g_{\alpha\beta}(ux)^3 + 4k^2(kq)g_{\alpha\beta}(ux)^3[u(1-x)]
\eeq
\[
- k^2(kq)g_{\alpha\beta}(ux)^3 - 3k^2(kq)g_{\alpha\beta}(ux)^2
[u(1-x)] + k^2(kq)g_{\alpha\beta}(ux)^2
\]
\[
- 2k^2k_\alpha k_\beta(ux)^4 + 2k^2k_\alpha
k_\beta(ux)^3 - 2k^2k_\alpha q_0g_{\beta 0}(ux)^3[u(1-x)]
\]
\[
- 2k^2k_\beta q_0g_{\alpha 0}(ux)^3[u(1-x)]
+ 2k^2q^2g_{\alpha\beta}(ux)^2[u(1-x)]^2- k^2q^2g_{\alpha\beta}(ux)^2[u(1-x)]
\]
\[
- k^2q^2g_{\alpha\beta}(ux)[u(1-x)]^2
+ 2k^2q^2g_{\alpha\beta}(ux)[u(1-x)]
- 2k^2q_0^2g_{\alpha\beta}(ux)^2[u(1-x)]^2
\]
\[
- 2k^2q_0^2g_{\alpha\beta}(ux)^2[u(1-x)]- 2k^2g_{\alpha\beta}m^2(ux)^2
+ k^2g_{\alpha\beta}m^2(ux)
\]
\[
+ 4(kq)^2g_{\alpha\beta}(ux)^2[u(1-x)]^2
- 2(kq)^2g_{\alpha\beta}(ux)^2[u(1-x)]
\]
\[
- 2(kq)^2g_{\alpha\beta}(ux)[u(1-x)]^2
- 4(kq)k_\alpha k_\beta(ux)^3[u(1-x)]
\]
\[
- 4(kq)k_\alpha q_0g_{\beta 0}(ux)^2[u(1-x)]^2
- 4(kq)k_\beta q_0g_{\alpha 0}(ux)^2[u(1-x)]^2
\]
\[
+ 4(kq)q^2g_{\alpha\beta}(ux)[u(1-x)]^3
- 3(kq)q^2g_{\alpha\beta}(ux)[u(1-x)]^2
- (kq)q^2g_{\alpha\beta}[u(1-x)]^3
\]
\[
+ (kq)q^2g_{\alpha\beta}[u(1-x)]^2
- 4(kq)q_0^2g_{\alpha\beta}(ux)[u(1-x)]^3
- 4(kq)g_{\alpha\beta}m^2(ux)[u(1-x)]
\]
\[
+ (kq)g_{\alpha\beta}m^2(ux)
+ (kq)g_{\alpha\beta}m^2[u(1-x)] + (kq)g_{\alpha\beta}m^2
\]
\[
 - 2k_\alpha k_\beta q^2(ux)^2[u(1-x)]^2
- 2k_\alpha k_\beta q^2(ux)[u(1-x)]^2
+ 8k_\alpha k_\beta q_0^2(ux)^2[u(1-x)]^2
\]
\[
 + 2k_\alpha k_\beta m^2(ux)^2
+ 2k_\alpha k_\beta m^2(ux)
 - 2k_\alpha q^2q_0g_{\beta 0}(ux)[u(1-x)]^3
\]
\[
+ 2k_\alpha q_0g_{\beta 0}m^2(ux)[u(1-x)]
- 2k_\beta q^2q_0g_{\alpha 0}(ux)[u(1-x)]^3
\]
\[
+ 2k_\beta q_0g_{\alpha 0}m^2(ux)[u(1-x)]
+ q^4 g_{\alpha\beta}[u(1-x)]^4
- q^4g_{\alpha\beta}[u(1-x)]^3
\]
\[
- 2q^2q_0^2g_{\alpha\beta}[u(1-x)]^4
+ 2q^2q_0^2g_{\alpha\beta}[u(1-x)]^3 - 2q^2g_{\alpha\beta}m^2[u(1-x)]^2
\]
\[
+ q^2g_{\alpha\beta}m^2[u(1-x)]
+ 2q_0^2g_{\alpha\beta}m^2[u(1-x)]^2
+ 2q_0^2g_{\alpha\beta}m^2[u(1-x)] + g_{\alpha\beta}m^4.
\]

Next we perform the Wick rotation and integration over the loop
momentum $p'$. The momentum integral diverges logarithmically. It is well known
that even this log of the ultraviolet cutoff cancels if one adds the
crossed diagram contribution.  A finite term breaking gauge invariance
remains as the only remnant of the divergence.  Since it is still
necessary to subtract the value of the light by light scattering tensor at zero
momentum in order to restore gauge invariance we will now perform the
subtraction directly in the ladder diagram contribution.

Integration is performed trivially and we obtain

\beq         \label{uvdiv}
{\tilde{\cal L}}_{\alpha\beta 00}=
\eeq
\[
-6\int_0^1dx\int_0^1duu(1-u)\{
\frac{1}{3}(\log{\frac{\Lambda^2}{\Delta}}-\frac{11}{6})
{\tilde L}_{\alpha\beta 00}^{(0)}
- \frac{1}{6\Delta}{\tilde L}_{\alpha\beta 00}^{(1)}
+\frac{1}{6\Delta^2}{\tilde L}_{\alpha\beta 00}^{(2)}\}.
\]

Next we do the subtraction. It is convenient first to perform auxiliary
integration by parts.  This gives one a chance to separate a subtraction term
connected  with the logarithm and to get rid of this log in the integrand
simultaneously. We use the identity

\beq
\int_0^1duu(1-u)\log{\frac{\Lambda^2}{\Delta}}=-\frac{(1-u)^2(2u+1)}{6}
\log{\frac{\Lambda^2}{\Delta}}|_0^1
\eeq
\[
-\int_0^1\frac{du}{6\Delta}(1-u)^2(2u+1)\{-k^2x(1-2xu)
\]
\[
-q^2(1-x)[1-2u(1-x)]+4(kq)ux(1-x)\}
\]
\[
=\frac{1}{6}\log{\frac{\Lambda^2}{m^2}}-\int_0^1\frac{du}{6\Delta}(1-u)^2
(2u+1)\{-k^2x(1-2xu)
\]
\[
-q^2(1-x)[1-2u(1-x)]+4(kq)ux(1-x)\}
\]
to deal with the log on the right hand side in \eq{uvdiv}.

Subtraction of the logarithmic term at zero momenta is performed simply by
omitting the log as well as the constant term after integration by parts.

Some finite terms in \eq{uvdiv} also need subtraction. As was mentioned
above it is necessary to subtract the value of the ladder contribution to
the light by light scattering tensor at zero external momentum. Considering
explicit expressions for the functions ${\tilde L}_{\alpha\beta 00}^{(i)}$  in
\eq{auxfunct} it is easy to see  that only terms $ 2g_{\alpha 0}g_{\beta
0}m^2 - 2g_{\alpha\beta}m^2$ in function ${\tilde L}_{\alpha\beta 00}^{(1)}$
and the term  $g_{\alpha\beta}m^4$ in function ${\tilde L}_{\alpha\beta
00}^{(2)}$ need subtraction. Now, subtraction is performed simply by the
replacements

\beq
\frac{1}{\Delta}\rightarrow \frac{1}{\Delta(k,q)}-\frac{1}{\Delta(0,0)}
\eeq
\[
=\frac{k^2xu(1-ux)+q^2u(1-x)[1-u(1-x)]-2(kq)u^2x(1-x)}{m^2\Delta},
\]
\[
\frac{1}{\Delta^2}\rightarrow
\frac{1}{\Delta^2(k,q)}-\frac{1}{\Delta^2(0,0)}
=(\frac{1}{\Delta(k,q)}-\frac{1}{\Delta(0,0)})
(\frac{1}{\Delta(k,q)}+\frac{1}{\Delta(0,0)})
\]
\[
=(2m^2-k^2xu(1-ux)-q^2u(1-x)[1-u(1-x)]
+2(kq)u^2x(1-x))
\]
\[
\frac{k^2xu(1-ux)+q^2u(1-x)[1-u(1-x)]
-2(kq)u^2x(1-x)}{m^4\Delta^2}
\]
in the denominators of the terms just discussed.

We obtain then the expression for the subtracted ladder diagram contribution

\beq         \label{ladder}
{\tilde{\cal L}}_{\alpha\beta 00}^{sub}=
\eeq
\[
-\int_0^1dx\int_0^1du\{\frac{1-u}{3\Delta}
{\tilde L}_{\alpha\beta 00}^{(1)sub}
+\frac{u(1-u)}{\Delta^2}{\tilde L}_{\alpha\beta 00}^{(2)sub}\},
\]

where

\beq
{\tilde L}_{\alpha\beta 00}^{(1)sub}=
- 2k^2g_{\alpha 0}g_{\beta 0}uvx(1-u) - 6k^2g_{\alpha 0}g_{\beta 0}uv(ux)
\eeq
\[
+2k^2g_{\alpha 0}g_{\beta 0}(ux)^2(1-u) + 6k^2g_{\alpha 0}g_{\beta 0}u(ux)^2
- k^2g_{\alpha 0}g_{\beta 0}vx(1-u)
\]
\[
+ k^2g_{\alpha 0}g_{\beta 0}x(ux)(1-u)+ 2k^2g_{\alpha\beta}uvx(1-u)
+ 6k^2g_{\alpha\beta}uv(ux)
\]
\[
- 2k^2g_{\alpha\beta}(ux)^2(1-u) - 12k^2g_{\alpha\beta}u(ux)^2
+ 3k^2g_{\alpha\beta}u(ux)+ k^2g_{\alpha\beta}vx(1-u)
\]
\[
- k^2g_{\alpha\beta}x(ux)(1-u)
+ 8(kq)g_{\alpha 0}g_{\beta 0}[u(1-x)](ux)(1-u)
\]
\[
+ 24(kq)g_{\alpha 0}g_{\beta 0}u(ux)[u(1-x)]
+ 4(kq)g_{\alpha 0}g_{\beta 0}(1-x)(ux)(1-u)
\]
\[
- 8(kq)g_{\alpha\beta}[u(1-x)](ux)(1-u)
- 36(kq)g_{\alpha\beta}u(ux)[u(1-x)]
\]
\[
+ 3(kq)g_{\alpha\beta}u(ux)+ 3(kq)g_{\alpha\beta}u[u(1-x)]
+ 3(kq)g_{\alpha\beta}u
\]
\[
- 4(kq)g_{\alpha\beta}(1-x)(ux)(1-u)+ 12k_\alpha k_\beta u(ux)^2
+ 6k_\alpha k_\beta u(ux)
\]
\[
- 2q^2g_{\alpha 0}g_{\beta 0}s[u(1-x)](1-u)
- 6q^2g_{\alpha 0}g_{\beta 0}su[u(1-x)]
\]
\[
- q^2g_{\alpha 0}g_{\beta 0}s(1-x)(1-u)
+ 2q^2g_{\alpha 0}g_{\beta 0}[u(1-x)][u(1-x)](1-u)
\]
\[
+ 6q^2g_{\alpha 0}g_{\beta 0}u[u(1-x)]^2
+ q^2g_{\alpha 0}g_{\beta 0}(1-x)[u(1-x)](1-u)
\]
\[
+ 2q^2g_{\alpha\beta}s[u(1-x)](1-u)
+ 6q^2g_{\alpha\beta}su[u(1-x)]
+ q^2g_{\alpha\beta}s(1-x)(1-u)
\]
\[
- 2q^2g_{\alpha\beta}[u(1-x)][u(1-x)](1-u)- 12q^2g_{\alpha\beta}u[u(1-x)]^2
+ 3q^2g_{\alpha\beta}u[u(1-x)]
\]
\[
- q^2g_{\alpha\beta}(1-x)[u(1-x)](1-u)
+ 12q_0^2g_{\alpha\beta}u[u(1-x)]^2
+ 6q_0^2g_{\alpha\beta}u[u(1-x)],
\]

\medskip

\beq
{\tilde L}_{\alpha\beta 00}^{(2)sub}=- k^4g_{\alpha\beta}v^2(ux)^2
+ k^4g_{\alpha\beta}(ux)^4- k^4g_{\alpha\beta}(ux)^3
\eeq
\[
+ 4k^2(kq)g_{\alpha\beta}v(ux)^2[u(1-x)] +
4k^2(kq)g_{\alpha\beta}(ux)^3[u(1-x)] -
k^2(kq)g_{\alpha\beta}(ux)^3
\]
\[
- 3k^2 (kq)g_{\alpha\beta}(ux)^2[u(1-x)] + k^2(kq)g_{\alpha\beta}(ux)^2
- 2k^2k_\alpha k_\beta(ux)^4+ 2k^2k_\alpha k_\beta(ux)^3
\]
\[
- 2k^2k_\alpha q_0g_{\beta 0}(ux)^3[u(1-x)] -
2k^2k_\beta q_0g_{\alpha 0}(ux)^3 [u(1-x)]
\]
\[
- 2k^2q^2g_{\alpha\beta}sv(ux)[u(1-x)]+ 2k^2q^2g_{\alpha\beta}
(ux)^2[u(1-x)]^2 - k^2q^2g_{\alpha\beta}(ux)^2[u(1-x)]
\]
\[
- k^2q^2g_{\alpha\beta}(ux)[u(1-x)]^2 + 2k^2q^2g_{\alpha\beta}(ux)[u(1-x)]
- 2k^2q_0^2g_{\alpha\beta}(ux)^2[u(1-x)]^2
\]
\[
- 2k^2q_0^2g_{\alpha\beta}(ux)^2[u(1-x)]
+ 2k^2g_{\alpha\beta}m^2v(ux) - 2k^2g_{\alpha\beta}m^2(ux)^2
+ k^2g_{\alpha\beta}m^2(ux)
\]
\[
- 2(kq)^2g_{\alpha\beta}(ux)^2[u(1-x)]
- 2(kq)^2g_{\alpha\beta}(ux)[u(1-x)]^2
\]
\[
- 4(kq)k_\alpha k_\beta(ux)^3[u(1-x)]
- 4(kq)k_\alpha q_0g_{\beta 0}(ux)^2[u(1-x)]^2
\]
\[
- 4(kq)k_\beta q_0g_{\alpha 0}(ux)^2[u(1-x)]^2
+ 4(kq)q^2g_{\alpha\beta}s(ux)[u(1-x)]^2
\]
\[
+ 4(kq)q^2g_{\alpha\beta}(ux)[u(1-x)]^3
- 3(kq)q^2g_{\alpha\beta}(ux)[u(1-x)]^2- (kq)q^2g_{\alpha\beta}[u(1-x)]^3
\]
\[
+ (kq)q^2g_{\alpha\beta}[u(1-x)]^2 - 4(kq)q_0^2g_{\alpha\beta}(ux)[u(1-x)]^3
- 8(kq)g_{\alpha\beta}m^2(ux)[u(1-x)]
\]
\[
+ (kq)g_{\alpha\beta}m^2(ux) + (kq)g_{\alpha\beta}m^2[u(1-x)]
+ (kq)g_{\alpha\beta}m^2 - 2k_\alpha k_\beta q^2(ux)^2[u(1-x)]^2
\]
\[
- 2k_\alpha k_\beta q^2(ux)[u(1-x)]^2
+ 8k_\alpha k_\beta q_0^2(ux)^2[u(1-x)]^2+ 2k_\alpha k_\beta m^2(ux)^2
\]
\[
+ 2k_\alpha k_\beta m^2(ux) - 2k_\alpha q^2q_0g_{\beta 0}(ux)[u(1-x)]^3
+ 2k_\alpha q_0g_{\beta 0}m^2 (ux)[u(1-x)]
\]
\[
- 2k_\beta q^2q_0g_{\alpha 0}(ux)[u(1-x)]^3
+ 2k_\beta q_0g_{\alpha 0}m^2(ux)[u(1-x)] - q^4g_{\alpha\beta}s^2[u(1-x)]^2
\]
\[
+ q^4g_{\alpha\beta}[u(1-x)]^4 - q^4g_{\alpha\beta}[u(1-x)]^3
- 2q^2q_0^2g_{\alpha\beta}[u(1-x)]^4
\]
\[
+ 2q^2q_0^2g_{\alpha\beta}[u(1-x)]^3 + 2 q^2g_{\alpha\beta}m^2s[u(1-x)]
- 2q^2g_{\alpha\beta}m^2[u(1-x)]^2
\]
\[
+ q^2g_{\alpha\beta}m^2[u(1-x)]
+ 2q_0^2g_{\alpha\beta}m^2[u(1-x)]^2
+ 2q_0^2g_{\alpha\beta}m^2[u(1-x)],
\]
\[
v=1-ux, \qquad s=1-u(1-x).
\]

\subsection{CROSSED DIAGRAM}

The symmetrized expression for the crossed diagram tensor in
\eq{croslght} has the form

\beq
{\tilde C}_{\alpha\beta 00}=-\frac{1}{D_1D_2D_3D_4}\{
2k^2p^2g_{\alpha 0}g_{\beta 0}- k^2p^2g_{\alpha\beta}
- 2k^2(pq)g_{\alpha 0}g_{\beta 0}
\eeq
\[
+ k^2(pq)g_{\alpha\beta} - 2k^2p_{0}p_{\alpha}g_{\beta 0}
- 2k^2p_{0}p_{\beta}g_{\alpha 0} + 2k^2p_{\alpha}p_{\beta}
+ k^2p_{\alpha}q_{0}g_{\beta 0} + k^2p_{\beta}q_{0}g_{\alpha 0}
\]
\[
- 2k^2g_{\alpha 0}g_{\beta 0}m^2 + k^2g_{\alpha\beta}m^2
- 2(kp)k_{\alpha}p_{\beta} - 2(kp)k_{\beta}p_{\alpha}
- 4(kp)p^2g_{\alpha 0}g_{\beta 0} + 2(kp)p^2g_{\alpha\beta}
\]
\[
+ 4(kp)(pq)g_{\alpha 0}g_{\beta 0} - 2(kp)(pq)g_{\alpha\beta}
+ 4(kp)p_{0}p_{\alpha}g_{\beta 0} + 4(kp)p_{0}p_{\beta}g_{\alpha 0}
- 2(kp)p_{\alpha}q_{0}g_{\beta 0}
\]
\[
- 2(kp)p_{\beta}q_{0}g_{\alpha 0} - 2(kp)q^2g_{\alpha 0}g_{\beta 0}
+ (kp)q^2g_{\alpha\beta} - 2(kp)q_{0}^2g_{\alpha\beta}
+ 4(kp)g_{\alpha 0}g_{\beta 0}m^2
\]
\[
- 2(kp)g_{\alpha\beta}m^2 + (kq)k_{\alpha}p_{\beta}
+ (kq)k_{\beta}p_{\alpha} + 2(kq)p^2g_{\alpha 0}g_{\beta 0}
- (kq)p^2g_{\alpha\beta} - 2(kq)p_{0}p_{\alpha}g_{\beta 0}
\]
\[
- 2(kq)p_{0}p_{\beta}g_{\alpha 0} + 2(kq)p_{0}q_{0}g_{\alpha\beta}
- 2(kq)g_{\alpha 0}g_{\beta 0}m^2 + (kq)g_{\alpha\beta}m^2
+ 2k_{\alpha}k_{\beta}p^2
\]
\[
- 2k_{\alpha}k_{\beta}(pq)- 2k_{\alpha}k_{\beta}m^2
+ 2k_{\alpha}p^2p_{0}g_{\beta 0} - k_{\alpha}p^2q_{0}g_{\beta 0}
- 2k_{\alpha}(pq)p_{0}g_{\beta 0} - 4k_{\alpha}p_{0}^2p_{\beta}
\]
\[
+ 4k_{\alpha}p_{0}p_{\beta}q_{0} + k_{\alpha}p_{0}q^2g_{\beta 0}
- 2k_{\alpha}p_{0}g_{\beta 0}m^2 + k_{\alpha}q_{0}g_{\beta 0}m^2
+ 2k_{\beta}p^2p_{0}g_{\alpha 0} - k_{\beta}p^2q_{0}g_{\alpha 0}
\]
\[
- 2k_{\beta}(pq)p_{0}g_{\alpha 0} - 4k_{\beta}p_{0}^2p_{\alpha}
+ 4k_{\beta}p_{0}p_{\alpha}q_{0} + k_{\beta}p_{0}q^2g_{\alpha 0}
- 2k_{\beta}p_{0}g_{\alpha 0}m^2 + k_{\beta}q_{0}g_{\alpha 0}m^2
\]
\[
+ 2p^4g_{\alpha 0}g_{\beta 0} - p^4g_{\alpha\beta}
- 4p^2(pq)g_{\alpha 0}g_{\beta 0} + 2p^2(pq)g_{\alpha\beta}
- 4p^2p_{0}p_{\alpha}g_{\beta 0} - 4p^2p_{0}p_{\beta}g_{\alpha 0}
\]
\[
+ 2p^2p_{\alpha}q_{0}g_{\beta 0} + 2p^2p_{\beta}q_{0}g_{\alpha 0}
+ 2p^2q^2g_{\alpha 0}g_{\beta 0} - p^2q^2g_{\alpha\beta}
+ 2p^2q_{0}^2g_{\alpha\beta} - 4p^2g_{\alpha 0}g_{\beta 0}m^2
\]
\[
+ 2p^2g_{\alpha\beta}m^2 + 4(pq)p_{0}p_{\alpha}g_{\beta 0}
+ 4(pq)p_{0}p_{\beta}g_{\alpha 0} - 4(pq)p_{0}q_{0}g_{\alpha\beta}
+ 4(pq)g_{\alpha 0}g_{\beta 0}m^2
\]
\[
- 2(pq)g_{\alpha\beta}m^2 + 8p_{0}^2p_{\alpha}p_{\beta}
+ 2p_{0}^2q^2g_{\alpha\beta} - 8p_{0}p_{\alpha}p_{\beta}q_{0}
- 2p_{0}p_{\alpha}q^2g_{\beta 0} + 4p_{0}p_{\alpha}g_{\beta 0}m^2
\]
\[
- 2p_{0}p_{\beta}q^2g_{\alpha 0} + 4p_{0}p_{\beta}g_{\alpha 0}m^2
- 2p_{\alpha}q_{0}g_{\beta 0}m^2 - 2p_{\beta}q_{0}g_{\alpha 0}m^2
- 2q^2g_{\alpha 0}g_{\beta 0}m^2
\]
\[
+ q^2g_{\alpha\beta}m^2 - 2q_{0}^2g_{\alpha\beta}m^2
+ 2g_{\alpha 0}g_{\beta 0}m^4 - g_{\alpha\beta}m^4\}.
\]

Next we combine denominators and perform momentum integration. Unlike the
case of the ladder diagram there are four different denominators now
and an extra Feynman parameter $z$ is needed to combine denominators

\beq
(1-z)D_4+z[(1-u)D_1+u[xD_2+(1-x)D_3]]=p''^2-\Delta_z,
\eeq
\[
p''=p-k[1-z(1-ux)]-q[1-z(1-u(1-x))],
\]
\[
\Delta_z=z\Delta-z(1-z)[k(1-ux)+q(1-u(1-x))]^2+(1-z)m^2,
\]
\[
\frac{1}{D_1D_2D_3D_4}=6\int_0^1dx\int_0^1du\int_0^1dzuz^2
\frac{1}{(p''^2-\Delta_z)^4}.
\]
Note that $\Delta_{|z=1}=\Delta$ and $p''_{|z=1}=p'$.

After the shift in the crossed diagram we have

\beq
{\tilde{\cal C}}_{\alpha\beta 00}=
-6\int_0^1dx\int_0^1du\int_0^1dzuz^2\int\frac{d^4p''}{\pi^2 i}\{
\frac{2}{3}p''^4{\tilde C}_{\alpha\beta 00}^{(0)}
\eeq
\[
+ \frac{1}{2}p''^2{\tilde C}_{\alpha\beta 00}^{(1)}
+ {\tilde C}_{\alpha\beta 00}^{(2)}\}
\frac{1}{(p''^2-\Delta_z)^4},
\]

where

\beq                \label{auxfunctcr}
{\tilde C}_{\alpha\beta 00}^{(0)}=- g_{\alpha 0}g_{\beta 0} +
g_{\alpha\beta}\equiv -{\tilde L}_{\alpha\beta 00}^{(0)},
\eeq

\medskip

\beq
{\tilde C}_{\alpha\beta 00}^{(1)}=8k^2g_{\alpha 0}g_{\beta 0}a^2
- 8k^2g_{\alpha 0}g_{\beta 0}a + 2k^2g_{\alpha 0}g_{\beta 0}
- 6k^2g_{\alpha\beta}a^2 + 6k^2g_{\alpha\beta}a
\eeq
\[
- k^2g_{\alpha\beta} + 16(kq)g_{\alpha 0} g_{\beta 0}ab
- 8(kq)g_{\alpha 0}g_{\beta 0}a - 8(kq)g_{\alpha 0}g_{\beta 0}b
+ 4 (kq)g_{\alpha 0}g_{\beta 0}
\]
\[
- 12(kq)g_{\alpha\beta}ab+ 6(kq)g_{\alpha\beta}a + 6(kq) g_{\alpha\beta}b
- 3(kq)g_{\alpha\beta} + 4k_{\alpha}k_{\beta}a^2 - 4k_{\alpha}k_{\beta}a
\]
\[
+ 2k_{\alpha}k_{\beta} - 8k_{\alpha}q_0g_{\beta 0}ab
+ 4k_{\alpha}q_0g_{\beta 0}a + 4k_{\alpha}q_0g_{\beta 0}b
- 2k_{\alpha}q_0g_{\beta 0} - 8k_{\beta}q_0g_{\alpha 0}ab
\]
\[
+ 4k_{\beta}q_0g_{\alpha 0}a + 4k_{\beta}q_0g_{\alpha 0}b
- 2k_{\beta}q_0g_{\alpha 0} + 8q^2g_{\alpha 0}g_{\beta 0}b^2
- 8q^2g_{\alpha 0}g_{\beta 0}b + 2q^2g_{\alpha 0}g_{\beta 0}
\]
\[
- 6q^2g_{\alpha\beta}b^2 + 6q^2g_{\alpha\beta}b - q^2g_{\alpha\beta}
+ 4q_0^2g_{\alpha\beta}b^2 - 4q_0^2g_{\alpha\beta}b + 2q_0^2g_{\alpha\beta}
- 4g_{\alpha 0}g_{\beta 0}m^2
\]
\[
+ 4g_{\alpha\beta}m^2,
\]

\medskip

\beq
{\tilde C}_{\alpha\beta 00}^{(2)}=2k^4g_{\alpha 0}g_{\beta 0}a^4
- 4k^4g_{\alpha 0}g_{\beta 0}a^3 + 2k^4g_{\alpha 0}g_{\beta 0}a^2
- k^4g_{\alpha\beta}a^4 + 2k^4g_{\alpha\beta}a^3
\eeq
\[
- k^4g_{\alpha\beta}a^2 + 8k^2(kq)g_{\alpha 0}g_{\beta 0}a^3b
- 4k^2(kq)g_{\alpha 0}g_{\beta 0}a^3 - 12k^2(kq)g_{\alpha 0}g_{\beta 0}a^2b
\]
\[
+ 6k^2(kq)g_{\alpha 0}g_{\beta 0}a^2 + 4k^2(kq)g_{\alpha 0}g_{\beta 0}ab
- 2k^2(kq)g_{\alpha 0}g_{\beta 0}a - 4k^2(kq)g_{\alpha\beta}a^3b
\]
\[
+ 2k^2(kq)g_{\alpha\beta}a^3 + 6k^2(kq)g_{\alpha\beta}a^2b
- 3k^2(kq)g_{\alpha\beta}a^2- 2k^2(kq)g_{\alpha\beta}ab
+ k^2(kq)g_{\alpha\beta}a
\]
\[
- 4k^2k_{\alpha}q_0g_{\beta 0}a^3b
+ 2k^2k_{\alpha}q_0g_{\beta 0}a^3 + 6k^2k_{\alpha}q_0g_{\beta 0}a^2b
- 3k^2k_{\alpha}q_0g_{\beta 0}a^2 - 2k^2k_{\alpha}q_0g_{\beta 0}
ab
\]
\[
+ k^2k_{\alpha}q_0g_{\beta 0}a - 4k^2k_{\beta}q_0g_{\alpha 0}a^3b
+ 2k^2k_{\beta}q_0g_{\alpha 0}a^3 + 6k^2k_{\beta}q_0g_{\alpha 0}a^2b
- 3k^2k_{\beta}q_0g_{\alpha 0}a^2
\]
\[
- 2k^2k_{\beta}q_0g_{\alpha 0}ab + k^2k_{\beta}q_0g_{\alpha 0}a
+ 4k^2q^2g_{\alpha 0}g_{\beta 0}a^2b^2 - 4k^2q^2g_{\alpha 0}g_{\beta 0}a^2b
\]
\[
+ 2k^2q^2g_{\alpha 0}g_{\beta 0}a^2 - 4k^2q^2g_{\alpha 0}g_{\beta 0}ab^2
+ 4k^2q^2g_{\alpha 0}g_{\beta 0}ab - 2k^2q^2g_{\alpha 0}g_{\beta 0}a
\]
\[
+ 2k^2q^2g_{\alpha 0}g_{\beta 0}b^2 - 2k^2q^2g_{\alpha 0}g_{\beta 0}b
- 2k^2q^2g_{\alpha\beta}a^2b^2 + 2k^2q^2g_{\alpha\beta}a^2b
- k^2q^2g_{\alpha\beta}a^2
\]
\[
+ 2k^2q^2g_{\alpha\beta}ab^2 - 2k^2q^2g_{\alpha\beta}ab
+ k^2q^2g_{\alpha\beta}a - k^2q^2g_{\alpha\beta}b^2
+ k^2q^2g_{\alpha\beta}b + 2k^2q_0^2g_{\alpha\beta}a^2
\]
\[
- 2k^2q_0^2g_{\alpha\beta}a - 4k^2g_{\alpha 0}g_{\beta 0}a^2m^2
+ 4k^2g_{\alpha 0}g_{\beta 0}am^2 - 2k^2g_{\alpha 0}g_{\beta 0}m^2
+ 2k^2g_{\alpha\beta}a^2m^2
\]
\[
- 2k^2g_{\alpha\beta}am^2 + k^2g_{\alpha\beta}m^2
+ 8(kq)^2g_{\alpha 0}g_{\beta 0}a^2b^2 - 8(kq)^2g_{\alpha 0}g_{\beta 0}a^2b
- 8(kq)^2g_{\alpha 0}g_{\beta 0}ab^2
\]
\[
+ 8(kq)^2g_{\alpha 0}g_{\beta 0}ab - 4(kq)^2g_{\alpha\beta}a^2b^2
+ 4(kq)^2g_{\alpha\beta}a^2b + 4(kq)^2g_{\alpha\beta}ab^2
- 4(kq)^2g_{\alpha\beta}ab
\]
\[
- 8(kq)k_{\alpha}q_0g_{\beta 0}a^2b^2 + 8(kq)k_{\alpha}q_0g_{\beta 0}a^2b
+ 8(kq)k_{\alpha}q_0g_{\beta 0}ab^2 - 8(kq)k_{\alpha}q_0g_{\beta 0}ab
\]
\[
- 8(kq)k_{\beta}q_0g_{\alpha 0}a^2b^2 + 8(kq)k_{\beta}q_0g_{\alpha 0}a^2b
+ 8(kq)k_{\beta}q_0g_{\alpha 0}ab^2 - 8(kq)k_{\beta}q_0g_{\alpha 0}ab
\]
\[
+ 8(kq)q^2g_{\alpha 0}g_{\beta 0}ab^3 - 12(kq)q^2g_{\alpha 0}g_{\beta 0}ab^2
+ 4(kq)q^2g_{\alpha 0}g_{\beta 0}ab - 4(kq)q^2g_{\alpha 0}g_{\beta 0}b^3
\]
\[
+ 6(kq)q^2g_{\alpha 0}g_{\beta 0}b^2 - 2(kq)q^2g_{\alpha 0}g_{\beta 0}b
- 4(kq)q^2g_{\alpha\beta}ab^3 + 6(kq)q^2g_{\alpha\beta}ab^2
\]
\[
- 2(kq)q^2g_{\alpha\beta}ab + 2(kq)q^2g_{\alpha\beta}b^3
- 3(kq)q^2g_{\alpha\beta}b^2 + (kq)q^2g_{\alpha\beta}b
- 8(kq)g_{\alpha 0}g_{\beta 0}abm^2
\]
\[
+ 4(kq)g_{\alpha 0}g_{\beta 0}am^2 + 4(kq)g_{\alpha 0}g_{\beta 0}bm^2
- 2(kq)g_{\alpha 0}g_{\beta 0}m^2 + 4(kq)g_{\alpha\beta}abm^2
\]
\[
- 2(kq)g_{\alpha\beta}am^2 - 2(kq)g_{\alpha\beta}bm^2
+ (kq)g_{\alpha\beta}m^2 + 2k_{\alpha}k_{\beta}q^2b^2
- 2k_{\alpha}k_{\beta}q^2b
\]
\[
+ 8k_{\alpha}k_{\beta}q_0^2a^2b^2 - 8k_{\alpha}k_{\beta}q_0^2a^2b
- 8k_{\alpha}k_{\beta}q_0^2ab^2 + 8k_{\alpha}k_{\beta}q_0^2ab
- 2k_{\alpha}k_{\beta}m^2
\]
\[
- 4k_{\alpha}q^2q_0g_{\beta 0}ab^3 + 6k_{\alpha}q^2q_0g_{\beta 0}ab^2
- 2k_{\alpha}q^2q_0g_{\beta 0}ab + 2k_{\alpha}q^2q_0g_{\beta 0}b^3
- 3k_{\alpha}q^2q_0g_{\beta 0}b^2
\]
\[
+ k_{\alpha}q^2q_0g_{\beta 0}b + 4k_{\alpha}q_0g_{\beta 0}abm^2
- 2k_{\alpha}q_0g_{\beta 0}am^2 - 2k_{\alpha}q_0g_{\beta 0}bm^2
+ k_{\alpha}q_0g_{\beta 0}m^2
\]
\[
- 4k_{\beta}q^2q_0g_{\alpha 0}ab^3 + 6k_{\beta}q^2q_0g_{\alpha 0}ab^2
- 2k_{\beta}q^2q_0g_{\alpha 0}ab + 2k_{\beta}q^2q_0g_{\alpha 0}b^3
- 3k_{\beta}q^2q_0g_{\alpha 0}b^2
\]
\[
+ k_{\beta}q^2q_0g_{\alpha 0}b + 4k_{\beta}q_0g_{\alpha 0}abm^2
- 2k_{\beta}q_0g_{\alpha 0}am^2 - 2k_{\beta}q_0g_{\alpha 0}bm^2
+ k_{\beta}q_0g_{\alpha 0}m^2
\]
\[
+ 2q^4g_{\alpha 0}g_{\beta 0}b^4 - 4q^4g_{\alpha 0}g_{\beta 0}b^3
+ 2q^4g_{\alpha 0}g_{\beta 0}b^2 - q^4g_{\alpha\beta}b^4
+ 2q^4g_{\alpha\beta}b^3 - q^4g_{\alpha\beta}b^2
\]
\[
- 4q^2g_{\alpha 0}g_{\beta 0}b^2m^2 + 4q^2g_{\alpha 0} g_{\beta 0}bm^2
- 2q^2g_{\alpha 0}g_{\beta 0}m^2 + 2q^2g_{\alpha\beta}b^2m ^2
- 2q^2g_{\alpha\beta}bm^2
\]
\[
+ q^2g_{\alpha\beta}m^2 - 2q_0^2g_{\alpha\beta} m^2
+ 2g_{\alpha 0}g_{\beta 0}m^4 - g_{\alpha\beta}m^4,
\]
\[
a=1-z(1-ux),
\]
\[
b=1-z[1-u(1-x)].
\]

Momentum integration leads to

\beq
{\tilde{\cal C}}_{\alpha\beta 00}=
\eeq
\[
-6\int_0^1dx\int_0^1du\int_0^1dzuz^2\{
\frac{2}{3}(\log{\frac{\Lambda^2}{\Delta}}-\frac{11}{6})
{\tilde C}_{\alpha\beta 00}^{(0)}
- \frac{1}{6\Delta_z}{\tilde C}_{\alpha\beta 00}^{(1)}
+ \frac{1}{6\Delta^2_z}{\tilde C}_{\alpha\beta 00}^{(2)}\}.
\]

Again we use integration by parts over $z$ to separate the ultraviolet
divergent term

\beq
\int_0^1dzz^2\log{\frac{\Lambda^2}{\Delta_z}}=-\frac{1-z^3}{3}
\log{\frac{\Lambda^2}{\Delta_z}}|_0^1
\eeq
\[
-\int_0^1\frac{dz}{3\Delta_z}(1-z^3)\{(\Delta-m^2)-(1-2z)
[k(1-ux)+q(1-u(1-x))]^2\}
\]
\[
=\frac{1}{3}\log{\frac{\Lambda^2}{m^2}}-\int_0^1\frac{dz}{3\Delta_z}(1-z^3)
\{(\Delta-m^2)-(1-2z)
[k(1-ux)+q(1-u(1-x))]^2\}.
\]

Subtraction in the logarithmic term at zero momenta is performed by
omitting the log as well as the constant term after integration by
parts, as
was done for the ladder diagram. Again this subtraction  does not
exhaust all necessary subtractions. The function
${\tilde C}_{\alpha\beta 00}^{(1)}$ contains terms
$(- 4g_{\alpha 0}g_{\beta 0}m^2 + 4g_{\alpha\beta}m^2)$ and function
${\tilde C}_{\alpha\beta 00}^{(2)}$  contains terms $2g_{\alpha 0}
g_{\beta 0}m^4 -g_{\alpha\beta}m^4$ which also require subtractions.

To perform subtractions this time it is sufficient to replace

\beq
\frac{1}{\Delta_z}\rightarrow
\frac{1}{\Delta_z(k,q)}-\frac{1}{\Delta_z(0,0)}
\eeq
\[
=\frac{z(m^2-\Delta)+z(1-z)[k(1-ux)+q(1-u(1-x))]^2}{m^2\Delta_z},
\]
\[
\frac{1}{\Delta_z^2}\rightarrow
\frac{1}{\Delta_z^2(k,q)}-\frac{1}{\Delta_z^2(0,0)}
=(\frac{1}{\Delta_z(k,q)}-\frac{1}{\Delta_z(0,0)})
(\frac{1}{\Delta_z(k,q)}+\frac{1}{\Delta_z(0,0)})
\]
\[
=(m^2+\Delta_z)
\frac{z(m^2-\Delta)+z(1-z)[k(1-ux)+q(1-u(1-x))]^2}{m^4\Delta_z^2}.
\]
in the denominators of the terms just discussed.

After subtraction we obtain

\beq \label{crtensor}
{\tilde{\cal C}}_{\alpha\beta 00}^{sub}=
\eeq
\[
-\int_0^1dx\int_0^1du\int_0^1dz\{
\frac{u}{3\Delta_z}{\tilde C}_{\alpha\beta 00}^{(1)sub}
+ \frac{uz^2}{\Delta^2_z}{\tilde C}_{\alpha\beta 00}^{(2)sub}\},
\]
where

\beq
{\tilde C}_{\alpha\beta 00}^{(1)sub}=-
24k^2g_{\alpha 0}g_{\beta 0}a^2z^2 + 24k^2g_{\alpha 0}g_{\beta 0}a z^2
\eeq
\[
+ 12k^2g_{\alpha 0}g_{\beta 0}v^2z^3(1-z)
+ 8k^2g_{\alpha 0} g_{\beta 0}vz^3(ux)
+ 4k^2g_{\alpha 0}g_{\beta 0}v(ux)
\]
\[
+ 4k^2g_{\alpha 0} g_{\beta 0}z^4(ux)^2
- 8k^2g_{\alpha 0}g_{\beta 0}z^4(ux)
+ 4k^2 g_{\alpha 0}g_{\beta 0}z^4 -
4k^2g_{\alpha 0}g_{\beta 0}z^3(ux)^2(1-z)
\]
\[
+ 8 k^2g_{\alpha 0}g_{\beta 0}z^3(ux)(1-z)
- 4k^2g_{\alpha 0}g_{\beta 0}z^3 (1-z)
- 6k^2g_{\alpha 0}g_{\beta 0}z^2 -
4k^2g_{\alpha 0}g_{\beta 0}z(ux)^ 2
\]
\[
+ 8k^2g_{\alpha 0}g_{\beta 0}z(ux) - 4k^2g_{\alpha 0}g_{\beta 0}z
+ 4k^2g_{\alpha 0}g_{\beta 0}(ux)^2(1-z)
- 8k^2g_{\alpha 0}g_{\beta 0}(ux)(1-z)
\]
\[
+ 4 k^2g_{\alpha 0}g_{\beta 0}(1-z) + 18k^2g_{\alpha\beta}a^2z^2
- 18k^2g_{\alpha\beta}az^2 - 12k^2g_{\alpha\beta}v^2z^3(1-z)
\]
\[
- 8k^2g_{\alpha\beta}vz^3(ux) - 4k^2g_{\alpha\beta}v(ux)
- 4k^2g_{\alpha\beta}z^4(ux)^2 + 8k^2g_{\alpha\beta}z^4(ux)
- 4k^2g_{\alpha\beta}z^4
\]
\[
+ 4k^2g_{\alpha\beta}z^3(ux)^2(1-z) - 8k^2g_{\alpha\beta}z^3(ux)(1-z)
+ 4k^2g_{\alpha\beta}z^3(1-z) + 3k^2g_{\alpha\beta}z^2
\]
\[
+ 4k^2g_{\alpha\beta}z(ux)^2 - 8k^2g_{\alpha\beta}z(ux)
+ 4k^2g_{\alpha\beta}z - 4k^2g_{\alpha\beta}(ux)^2(1-z)
\]
\[
+ 8k^2g_{\alpha\beta}(ux)(1-z) - 4k^2g_{\alpha\beta}(1-z)
- 48(kq)g_{\alpha 0}g_{\beta 0}abz^2
+ 24(kq)g_{\alpha 0}g_{\beta 0}az^2
\]
\[
+ 24(kq)g_{\alpha 0}g_{\beta 0}bz^2 + 24(kq)g_{\alpha 0}g_{\beta 0}svz^3(1-z)
+ 8(kq)g_{\alpha 0}g_{\beta 0}z^4(ux)[u(1-x)]
\]
\[
- 8(kq)g_{\alpha 0}g_{\beta 0}z^4(ux)
- 8(kq)g_{\alpha 0}g_{\beta 0}z^4[u(1-x)]
+ 8(kq)g_{\alpha 0}g_{\beta 0}z^4
\]
\[
- 8(kq)g_{\alpha 0}g_{\beta 0}z^3(ux)[u(1-x)](1-z)
- 16(kq)g_{\alpha 0}g_{\beta 0}z^3(ux)[u(1-x)]
\]
\[
+ 8(kq)g_{\alpha 0}g_{\beta 0}z^3(ux)(1-z)
+ 8(kq)g_{\alpha 0}g_{\beta 0}z^3[u(1-x)](1-z)
\]
\[
- 8(kq)g_{\alpha 0}g_{\beta 0}z^3(1-z)
- 12(kq)g_{\alpha 0}g_{\beta 0}z^2
- 8(kq)g_{\alpha 0}g_{\beta 0}z(ux)[u(1-x)]
\]
\[
+ 8(kq)g_{\alpha 0}g_{\beta 0}z(ux) + 8(kq)g_{\alpha 0}g_{\beta 0}z[u(1-x)]
 - 8(kq)g_{\alpha 0}g_{\beta 0}z
\]
\[
+ 8(kq)g_{\alpha 0}g_{\beta 0}(ux)[u(1-x)](1-z)
- 8(kq)g_{\alpha 0}g_{\beta 0}(ux)[u(1-x)]
\]
\[
- 8(kq)g_{\alpha 0}g_{\beta 0}(ux)(1-z)
- 8(kq)g_{\alpha 0}g_{\beta 0}[u(1-x)](1-z)
+ 8(kq)g_{\alpha 0}g_{\beta 0}(1-z)
\]
\[
+ 36(kq)g_{\alpha\beta}abz^2
- 18(kq)g_{\alpha\beta}az^2
- 18(kq)g_{\alpha\beta}bz^2 - 24(kq)g_{\alpha\beta}svz^3(1-z)
\]
\[
- 8(kq)g_{\alpha\beta}z^4(ux)[u(1-x)]
+ 8(kq)g_{\alpha\beta}z^4(ux) + 8(kq)g_{\alpha\beta}z^4[u(1-x)]
\]
\[
-8(kq)g_{\alpha\beta}z^4
+ 8(kq)g_{\alpha\beta}z^3(ux)[u(1-x)](1-z)
+ 16(kq)g_{\alpha\beta}z^3(ux)[u(1-x)]
\]
\[
- 8(kq)g_{\alpha\beta}z^3(ux)(1-z) - 8(kq)g_{\alpha\beta}z^3[u(1-x)](1-z)
+ 8(kq)g_{\alpha\beta}z^3(1-z)
\]
\[
+ 9(kq)g_{\alpha\beta}z^2 + 8(kq)g_{\alpha\beta}z(ux)[u(1-x)]
- 8(kq)g_{\alpha\beta}z(ux) - 8(kq)g_{\alpha\beta}z[u(1-x)]
\]
\[
+ 8(kq)g_{\alpha\beta}z - 8(kq)g_{\alpha\beta}(ux)[u(1-x)](1-z)
+ 8(kq)g_{\alpha\beta}(ux)[u(1-x)]
\]
\[
+ 8(kq)g_{\alpha\beta}(ux)(1-z) + 8(kq)g_{\alpha\beta}[u(1-x)](1-z)
- 8(kq)g_{\alpha\beta}(1-z) - 12k_{\alpha}k_{\beta}a^2z^2
\]
\[
+ 12k_{\alpha}k_{\beta}az^2 - 6k_{\alpha}k_{\beta}z^2
+ 24k_{\alpha}q_0g_{\beta 0}abz^2 - 12k_{\alpha}q_0g_{\beta 0}az^2
- 12k_{\alpha}q_0g_{\beta 0}bz^2
\]
\[
+ 6k_{\alpha}q_0g_{\beta 0}z^2 + 24k_{\beta}q_0g_{\alpha 0}abz^2
- 12k_{\beta}q_0g_{\alpha 0}az^2 - 12k_{\beta}q_0g_{\alpha 0}bz^2
+ 6k_{\beta}q_0g_{\alpha 0}z^2
\]
\[
- 24q^2g_{\alpha 0}g_{\beta 0}b^2z^2 + 24q^2g_{\alpha 0}g_{\beta 0}bz^2
+ 12q^2g_{\alpha 0}g_{\beta 0}s^2z^3(1-z)
\]
\[
+ 8q^2g_{\alpha 0}g_{\beta 0}sz^3[u(1-x)]
+ 4q^2g_{\alpha 0}g_{\beta 0}s[u(1-x)]
+ 4q^2g_{\alpha 0}g_{\beta 0}z^4[u(1-x)]^2
\]
\[
- 8q^2g_{\alpha 0}g_{\beta 0}z^4[u(1-x)]  + 4q^2g_{\alpha 0}g_{\beta 0}z^4
- 4q^2g_{\alpha 0}g_{\beta 0}z^3[u(1-x)]^2(1-z)
\]
\[
+ 8q^2g_{\alpha 0}g_{\beta 0}z^3[u(1-x)](1-z)
- 4q^2g_{\alpha 0}g_{\beta 0}z^3(1-z)
- 6q^2g_{\alpha 0}g_{\beta 0}z^2
\]
\[
- 4q^2g_{\alpha 0}g_{\beta 0}z[u(1-x)]^2
+ 8q^2g_{\alpha 0}g_{\beta 0}z[u(1-x)] - 4q^2g_{\alpha 0}g_{\beta 0}z
\]
\[
+ 4q^2g_{\alpha 0}g_{\beta 0}[u(1-x)]^2(1-z)
- 8q^2g_{\alpha 0}g_{\beta 0}[u(1-x)](1-z)
+ 4q^2g_{\alpha 0}g_{\beta 0}(1-z)
\]
\[
+ 18q^2g_{\alpha\beta}b^2z^2 - 18q^2g_{\alpha\beta}bz^2
- 12q^2g_{\alpha\beta}s^2z^3(1-z) - 8q^2g_{\alpha\beta}sz^3[u(1-x)]
\]
\[
- 4q^2g_{\alpha\beta}s[u(1-x)] - 4q^2g_{\alpha\beta}z^4[u(1-x)]^2
+ 8q^2g_{\alpha\beta}z^4[u(1-x)] - 4q^2g_{\alpha\beta}z^4
\]
\[
+ 4q^2g_{\alpha\beta}z^3[u(1-x)]^2(1-z)
- 8q^2g_{\alpha\beta}z^3[u(1-x)](1-z) + 4q^2g_{\alpha\beta}z^3(1-z)
\]
\[
+ 3q^2g_{\alpha\beta}z^2 + 4q^2g_{\alpha\beta}z[u(1-x)]^2
- 8q^2g_{\alpha\beta}z[u(1-x)] + 4q^2g_{\alpha\beta}z
\]
\[
- 4q^2g_{\alpha\beta}[u(1-x)]^2(1-z) + 8q^2g_{\alpha\beta}[u(1-x)](1-z)
- 4q^2g_{\alpha\beta}(1-z) - 12q_0^2g_{\alpha\beta}b^2z^2
\]
\[
+ 12q_0^2g_{\alpha\beta}bz^2 - 6q_0^2g_{\alpha\beta}z^2,
\]

\medskip

\beq
{\tilde C}_{\alpha\beta 00}^{(2)sub}=2k^4g_{\alpha 0}g_{\beta 0}a^4
- 4k^4g_{\alpha 0}g_{\beta 0}a^3 + 2k^4g_{\alpha 0}g_{\beta 0}a^2
- 2k^4g_{\alpha 0}g_{\beta 0}v^4z^2(1-z)^2
\eeq
\[
- 4k^4g_{\alpha 0}g_{\beta 0}v^3z^2(ux)(1-z)
- 2k^4g_{\alpha 0}g_{\beta 0}v^2z^2(ux)^2
- k^4g_{\alpha\beta}a^4 + 2k^4g_{\alpha\beta}a^3 - k^4g_{\alpha\beta}a^2
\]
\[
+ k^4g_{\alpha\beta}v^4z^2(1-z)^2 + 2k^4g_{\alpha\beta}v^ 3z^2(ux)(1-z)
+ k^4g_{\alpha\beta}v^2z^2(ux)^2
\]
\[
+ 8k^2(kq)g_{\alpha 0}g_{\beta 0}a^3b - 4k^2(kq)g_{\alpha 0}g_{\beta 0}a^3
- 12k^2(kq)g_{\alpha 0}g_{\beta 0}a^2b + 6k^2(kq)g_{\alpha 0}g_{\beta 0}a^2
\]
\[
+ 4k^2(kq)g_{\alpha 0}g_{\beta 0}ab
- 2k^2(kq)g_{\alpha 0}g_{\beta 0}a
- 8k^2(kq)g_{\alpha 0}g_{\beta 0}sv^3z^2(1-z)^2
\]
\[
- 8k^2(kq)g_{\alpha 0}g_{\beta 0}sv^2z^2(ux)(1-z)
+ 8k^2(kq)g_{\alpha 0}g_{\beta 0}v^2z^2(ux)[u(1-x)](1-z)
\]
\[
+ 8k^2(kq)g_{\alpha 0}g_{\beta 0}vz^2(ux)^2[u(1-x)]
- 4k^2(kq)g_{\alpha\beta}a^3b + 2k^2(kq)g_{\alpha\beta}a^3
\]
\[
+ 6k^2(kq)g_{\alpha\beta}a^2b- 3k^2(kq)g_{\alpha\beta}a^2
- 2k^2(kq)g_{\alpha\beta}ab + k^2(kq)g_{\alpha\beta}a
\]
\[
+ 4k^2(kq)g_{\alpha\beta}sv^3z^2(1-z)^2
+ 4k^2(kq)g_{\alpha\beta}sv^2z^2(ux)(1-z)
\]
\[
- 4k^2(kq)g_{\alpha\beta}v^2z^2(ux)[u(1-x)](1-z)
- 4k^2(kq)g_{\alpha\beta}vz^2(ux)^2[u(1-x)]
\]
\[
- 4k^2k_{\alpha}q_0g_{\beta 0}a^3b + 2k^2k_{\alpha}q_0g_{\beta 0}a^3
+ 6k^2k_{\alpha}q_0g_{\beta 0}a^2b - 3k^2k_{\alpha}q_0g_{\beta 0}a^2
- 2k^2k_{\alpha}q_0g_{\beta 0}ab
\]
\[
+ k^2k_{\alpha}q_0g_{\beta 0}a - 4k^2k_{\beta}q_0g_{\alpha 0}a^3b
+ 2k^2k_{\beta}q_0g_{\alpha 0}a^3 + 6k^2k_{\beta}q_0g_{\alpha 0}a^2b
- 3k^2k_{\beta}q_0g_{\alpha 0}a^2
\]
\[
- 2k^2k_{\beta}q_0g_{\alpha 0}ab+ k^2k_{\beta}q_0g_{\alpha 0}a
+ 4k^2q^2g_{\alpha 0}g_{\beta 0}a^2b^2 - 4k^2q^2g_{\alpha 0}g_{\beta 0}a^2b
\]
\[
+ 2k^2q^2g_{\alpha 0}g_{\beta 0}a^2- 4k^2q^2g_{\alpha 0}g_{\beta 0}ab^2
+ 4k^2q^2g_{\alpha 0}g_{\beta 0}ab - 2k^2q^2g_{\alpha 0}g_{\beta 0}a
\]
\[
+ 2k^2q^2g_{\alpha 0}g_{\beta 0}b^2 - 2k^2q^2g_{\alpha 0}g_{\beta 0}b
- 4k^2q^2g_{\alpha 0}g_{\beta 0}s^2v^2z^2(1-z)^2
\]
\[
- 4k^2q^2g_{\alpha 0}g_{\beta 0}s^2vz^2(ux)(1-z)
- 4k^2q^2g_{\alpha 0}g_{\beta 0}sv^2z^2[u(1-x)](1-z)
\]
\[
- 4k^2q^2g_{\alpha 0}g_{\beta 0}svz^2(ux)[u(1-x)]
- 2k^2q^2g_{\alpha\beta}a^2b^2 + 2k^2q^2g_{\alpha\beta}a^2b
- k^2q^2g_{\alpha\beta}a^2
\]
\[
+ 2k^2q^2g_{\alpha\beta}ab^2 - 2k^2q^2g_{\alpha\beta}ab
+ k^2q^2g_{\alpha\beta}a- k^2q^2g_{\alpha\beta}b^2
+ k^2q^2g_{\alpha\beta}b
\]
\[
 + 2k^2q^2g_{\alpha\beta}s^2v^2z^2(1-z)^2
+ 2k^2q^2g_{\alpha\beta}s^2vz^2(ux)(1-z)
\]
\[
+ 2k^2q^2g_{\alpha\beta}sv^2z^2[u(1-x)](1-z)
+ 2k^2q^2g_{\alpha\beta}svz^2(ux)[u(1-x)] + 2k^2q_0^2g_{\alpha\beta}a^2
\]
\[
- 2k^2q_0^2g_{\alpha\beta}a - 4k^2g_{\alpha 0}g_{\beta 0}a^2m^2
+ 4k^2g_{\alpha 0}g_{\beta 0}am^2 + 2k^2g_{\alpha 0}g_{\beta 0}m^2v^2z^2(1-z)
\]
\[
+ 2k^2g_{\alpha 0}g_{\beta 0}m^2v^2z(1-z)^2
+ 2k^2g_{\alpha 0}g_{\beta 0}m^2v^2z(1-z)
+ 2k^2g_{\alpha 0}g_{\beta 0}m^2vz^2(ux)
\]
\[
+ 2k^2g_{\alpha 0}g_{\beta 0}m^2vz(ux)(1-z)
+ 2k^2g_{\alpha 0}g_{\beta 0}m^2vz(ux) - 2k^2g_{\alpha 0}g_{\beta 0}m^2
+ 2k^2g_{\alpha\beta}a^2m^2
\]
\[
- 2k^2g_{\alpha\beta}am^2 - k^2g_{\alpha\beta}m^2v^2z^2(1-z)
- k^2g_{\alpha\beta}m^2v^2z(1-z)^2
- k^2g_{\alpha\beta}m^2v^2z(1-z)
\]
\[
- k^2g_{\alpha\beta}m^2vz^2(ux) - k^2g_{\alpha\beta}m^2vz(ux)(1-z)
- k^2g_{\alpha\beta}m^2vz(ux) + k^2g_{\alpha\beta}m^2
\]
\[
+ 8(kq)^2g_{\alpha 0}g_{\beta 0}a^2b^2 - 8(kq)^2g_{\alpha 0}g_{\beta 0}a^2b
- 8(kq)^2g_{\alpha 0}g_{\beta 0}ab^2 + 8(kq)^2g_{\alpha 0}g_{\beta 0}ab
\]
\[
- 8(kq)^2g_{\alpha 0}g_{\beta 0}s^2v^2z^2(1-z)^2
+ 16(kq)^2g_{\alpha 0}g_{\beta 0}svz^2(ux)[u(1-x)](1-z)
\]
\[
- 8(kq)^2g_{\alpha 0}g_{\beta 0}z^2(ux)^2[u(1-x)]^2
- 4(kq)^2g_{\alpha\beta}a^2b^2 + 4(kq)^2g_{\alpha\beta}a^2b
+ 4(kq)^2g_{\alpha\beta}ab^2
\]
\[
- 4(kq)^2g_{\alpha\beta}ab + 4(kq)^2g_{\alpha\beta}s^2v^2z^2(1-z)^2
- 8(kq)^2g_{\alpha\beta}svz^2(ux)[u(1-x)](1-z)
\]
\[
+ 4(kq)^2g_{\alpha\beta}z^2(ux)^2[u(1-x)]^2
- 8(kq)k_{\alpha}q_0g_{\beta 0}a^2b^2 + 8(kq)k_{\alpha}q_0g_{\beta 0}a^2b
\]
\[
+ 8(kq)k_{\alpha}q_0g_{\beta 0}ab^2 - 8(kq)k_{\alpha}q_0g_{\beta 0}ab
- 8(kq)k_{\beta}q_0g_{\alpha 0}a^2b^2
+ 8(kq)k_{\beta}q_0g_{\alpha 0}a^2b
\]
\[
+ 8(kq)k_{\beta}q_0g_{\alpha 0}ab^2 - 8(kq)k_{\beta}q_0g_{\alpha 0}ab
+ 8(kq)q^2g_{\alpha 0}g_{\beta 0}ab^3 - 12(kq)q^2g_{\alpha 0}g_{\beta 0}ab^2
\]
\[
+ 4(kq)q^2g_{\alpha 0}g_{\beta 0}ab
- 4(kq)q^2g_{\alpha 0}g_{\beta 0}b^3 + 6(kq)q^2g_{\alpha 0}g_{\beta 0}b^2
- 2(kq)q^2g_{\alpha 0}g_{\beta 0}b
\]
\[
- 8(kq)q^2g_{\alpha 0}g_{\beta 0}s^3vz^2(1-z)^2
- 8(kq)q^2g_{\alpha 0}g_{\beta 0}s^2vz^2[u(1-x)](1-z)
\]
\[
+ 8(kq)q^2g_{\alpha 0}g_{\beta 0}s^2z^2(ux)[u(1-x)](1-z)
+ 8(kq)q^2g_{\alpha 0}g_{\beta 0}sz^2(ux)[u(1-x)]^2
\]
\[
- 4(kq)q^2g_{\alpha\beta}ab^3 + 6(kq)q^2g_{\alpha\beta}ab^2
- 2(kq)q^2g_{\alpha\beta}ab + 2(kq)q^2g_{\alpha\beta}b^3
- 3(kq)q^2g_{\alpha\beta}b^2
\]
\[
+ (kq)q^2g_{\alpha\beta}b  + 4(kq)q^2g_{\alpha\beta}s^3vz^2(1-z)^2
+ 4(kq)q^2g_{\alpha\beta}s^2vz^2[u(1-x)](1-z)
\]
\[
- 4(kq)q^2g_{\alpha\beta}s^2z^2(ux)[u(1-x)](1-z)
- 4(kq)q^2g_{\alpha\beta}sz^2(ux)[u(1-x)]^2
\]
\[
- 8(kq)g_{\alpha 0}g_{\beta 0}abm^2
+ 4(kq)g_{\alpha 0}g_{\beta 0}am^2 + 4(kq)g_{\alpha 0}g_{\beta 0}bm^2
\]
\[
+ 4(kq)g_{\alpha 0}g_{\beta 0}m^2svz^2(1-z)
+ 4(kq)g_{\alpha 0}g_{\beta 0}m^2svz(1-z)^2
\]
\[
+ 4(kq)g_{\alpha 0}g_{\beta 0}m^2svz(1-z)
- 4(kq)g_{\alpha 0}g_{\beta 0}m^2z^2(ux)[u(1-x)]
\]
\[
- 4(kq)g_{\alpha 0}g_{\beta 0}m^2z(ux)[u(1-x)](1-z)
- 4(kq)g_{\alpha 0}g_{\beta 0}m^2z(ux)[u(1-x)]
\]
\[
- 2(kq)g_{\alpha 0}g_{\beta 0}m^2 + 4(kq)g_{\alpha\beta}abm^2
- 2(kq)g_{\alpha\beta}am^2 - 2(kq)g_{\alpha\beta}bm^2
\]
\[
- 2(kq)g_{\alpha\beta}m^2svz^2(1-z) - 2(kq)g_{\alpha\beta}m^2svz(1-z)^2
- 2(kq)g_{\alpha\beta}m^2svz(1-z)
\]
\[
+ 2(kq)g_{\alpha\beta}m^2z^2(ux)[u(1-x)]
+ 2(kq)g_{\alpha\beta}m^2z(ux)[u(1-x)](1-z)
\]
\[
+ 2(kq)g_{\alpha\beta}m^2z(ux)[u(1-x)]
+ (kq)g_{\alpha\beta}m^2 + 2k_{\alpha}k_{\beta}q^2b^2
- 2k_{\alpha}k_{\beta}q^2b + 8k_{\alpha}k_{\beta}q_0^2a^2b^2
\]
\[
- 8k_{\alpha}k_{\beta}q_0^2a^2b - 8k_{\alpha}k_{\beta}q_0^2ab^2
+ 8k_{\alpha}k_{\beta}q_0^2ab - 2k_{\alpha}k_{\beta}m^2
- 4k_{\alpha}q^2q_0g_{\beta 0}ab^3
\]
\[
+ 6k_{\alpha}q^2q_0g_{\beta 0}ab^2 - 2k_{\alpha}q^2q_0g_{\beta 0}ab
+ 2k_{\alpha}q^2q_0g_{\beta 0}b^3 - 3k_{\alpha}q^2q_0g_{\beta 0}b^2
+ k_{\alpha}q^2q_0g_{\beta 0}b
\]
\[
+ 4k_{\alpha}q_0g_{\beta 0}abm^2 - 2k_{\alpha}q_0g_{\beta 0}am^2
- 2k_{\alpha}q_0g_{\beta 0}bm^2 + k_{\alpha}q_0g_{\beta 0}m^2
- 4k_{\beta}q^2q_0g_{\alpha 0}ab^3
\]
\[
+ 6k_{\beta}q^2q_0g_{\alpha 0}ab^2 - 2k_{\beta}q^2q_0g_{\alpha 0}ab
+ 2k_{\beta}q^2q_0g_{\alpha 0}b^3 - 3k_{\beta}q^2q_0g_{\alpha 0}b^2
+ k_{\beta}q^2q_0g_{\alpha 0}b
\]
\[
+ 4k_{\beta}q_0g_{\alpha 0}abm^2 - 2k_{\beta}q_0g_{\alpha 0}am^2
- 2k_{\beta}q_0g_{\alpha 0}bm^2 + k_{\beta}q_0g_{\alpha 0}m^2
+ 2q^4g_{\alpha 0}g_{\beta 0}b^4
\]
\[
- 4q^4g_{\alpha 0}g_{\beta 0}b^3 + 2q^4g_{\alpha 0}g_{\beta 0}b^2
- 2q^4g_{\alpha 0}g_{\beta 0}s^4z^2(1-z)^2
\]
\[
- 4q^4g_{\alpha 0}g_{\beta 0}s^3z^2[u(1-x)](1-z)
- 2q^4g_{\alpha 0}g_{\beta 0}s^2z^2[u(1-x)]^2 - q^4g_{\alpha\beta}b^4
+ 2q^4g_{\alpha\beta}b^3
\]
\[
- q^4g_{\alpha\beta}b^2 + q^4g_{\alpha\beta}s^4z^2(1-z)^2
+ 2q^4g_{\alpha\beta}s^3z^2[u(1-x)](1-z) + q^4g_{\alpha\beta}s^2z^2[u(1-x)]^2
\]
\[
- 4q^2g_{\alpha 0}g_{\beta 0}b^2m^2 + 4q^2g_{\alpha 0}g_{\beta 0}bm^2
+ 2q^2g_{\alpha 0}g_{\beta 0}m^2s^2z^2(1-z)
\]
\[
+ 2q^2g_{\alpha 0}g_{\beta 0}m^2s^2z(1-z)^2
+ 2q^2g_{\alpha 0}g_{\beta 0}m^2s^2z(1-z)
+ 2q^2g_{\alpha 0}g_{\beta 0}m^2sz^2[u(1-x)]
\]
\[
+ 2q^2g_{\alpha 0}g_{\beta 0}m^2sz[u(1-x)](1-z)
+ 2q^2g_{\alpha 0}g_{\beta 0}m^2sz[u(1-x)]
- 2q^2g_{\alpha 0}g_{\beta 0}m^2
\]
\[
+ 2q^2g_{\alpha\beta}b^2m^2 - 2q^2g_{\alpha\beta}bm^2
- q^2g_{\alpha\beta}m^2s^2z^2(1-z) - q^2g_{\alpha\beta}m^2s^2z(1-z)^2
\]
\[
- q^2g_{\alpha\beta}m^2s^2z(1-z) - q^2g_{\alpha\beta}m^2sz^2[u(1-x)]
- q^2g_{\alpha\beta}m^2sz[u(1-x)](1-z)
\]
\[
- q^2g_{\alpha\beta}m^2sz[u(1-x)]+ q^2g_{\alpha\beta}m^2
- 2q_0^2g_{\alpha\beta}m^2.
\]

\subsection{LOW FREQUENCY LIMIT}

We have obtained in \eq{ladder} and \eq{crtensor} explicit expressions
for the contributions to the light by light scattering tensor induced by
the ladder and crossed diagrams, respectively. These expressions have
the form of the two- and three-dimensional integrals over the Feynman
parameters. A very important check of the correctness of all preceding
calculations may be performed now. Namely, we are going to compare the
low-frequency limit of the expressions above for the light by
light scattering tensor with the well known expression given by the
Euler-Heisenberg Lagrangian. Leading terms in the expansion of the tensor in
\eq{lghtbylght} over $k$ and $q$ should coincide with the Euler-Heisenberg
Lagrangian. Naively, it is  far from evident that the expansion of the
integrals over Feynman parameters obtained above  will start with
terms at least quartic in momenta and that these leading  terms will be at
least quadratic simultaneously in $q$ and $k$.

Performing the low frequency expansion in \eq{ladder} and \eq{crtensor} we
obtain

\beq
{\cal S}_{\alpha\beta 00}^{low}\equiv \lim_{k->0,q->0}
[2{\tilde{\cal L}}_{\alpha\beta 00}^{sub}
+{\tilde{\cal C}}_{\alpha\beta 00}^{sub}]
\eeq
\[
=\frac{13}{315m^4}k^2(kq)g_{\beta 0}g_{\alpha 0}
+ \frac{38}{315m^4}k^2(kq)g_{\alpha\beta} +
\frac{1}{126m^4}k^2k_{\beta}q_{0}g_{\alpha 0}
\]
\[
+ \frac{1}{126m^4}k^2k_{\alpha}q_{0}g_{\beta 0}
- \frac{7}{45m^4}k^2q^2g_{\beta 0}g_{\alpha 0}
+ \frac{1}{9m^4}k^2q^2g_{\alpha\beta}
- \frac{7}{45m^4} k^2q_{0}^2g_{\alpha\beta}
\]
\[
+ \frac{1}{15m^4}(kq)^2g_{\beta 0}g_{\alpha 0}
- \frac{7}{45m^4}(kq)^2g_{\alpha\beta}
- \frac{4}{35m^4}(kq)k_{\beta}k_{\alpha}
-\frac{1}{15m^4}(kq)k_{\beta}q_{0} g_{\alpha 0}
\]
\[
- \frac{1}{15m^4}(kq)k_{\alpha}q_{0}g_{\beta 0}
+ \frac{13}{315m^4}(kq)q^2g_{\beta 0}g_{\alpha 0}
+ \frac{38}{315m^4}(kq)q^2g_{\alpha\beta}
- \frac{4}{35m^4}(kq)q_{0}^2 g_{\alpha\beta}
\]
\[
+ \frac{1}{5m^4}(kq)g_{\beta 0}g_{\alpha 0}m^2
+ \frac{3}{5m^4}(kq)g_{\alpha\beta}m^2
- \frac{7}{45m^4}k_{\beta}k_{\alpha}q^2
+ \frac{1}{15m^4}k_{\beta}k_{\alpha}q_{0}^2
\]
\[
+ \frac{1}{126m^4}k_{\beta}q^2q_{0}g_{\alpha 0}
+ \frac{1}{30m^4}k_{\beta}q_{0}g_{\alpha 0}m^2
+ \frac{1}{126m^4}k_{\alpha}q^2q_{0}g_{\beta 0}
+ \frac{1}{30m^4}k_{\alpha}q_{0}g_{\beta 0}m^2.
\]

Note that some undesirable terms containing too small a number of factors
of $k$ or $q$
emerged in the expansion. One should not be too disappointed at this stage
since we have performed above a transformation of the light by light
scattering tensor taking into account that further integration
over $k$ kills all odd factors in $k$. It  is  easy to see that all
unwanted
terms are  odd in $k$ and, hence, they disappear after the integration over
$k$.

Really, it is not difficult to calculate second ladder diagram for the light
by light scattering separately and to obtain a low frequency asymptote
which does not contain unwanted terms at all

\beq
\tilde{{\cal S}}_{\alpha\beta 00}^{low}\equiv \lim_{k->0,q->0}
[{\tilde{\cal L}}_{\alpha\beta 00}^{1sub}
+{\tilde{\cal L}}_{\alpha\beta 00}^{2sub}
+{\tilde{\cal C}}_{\alpha\beta 00}^{sub}]
\eeq
\[
= - \frac{7}{45m^4}k^2q^2g_{\beta 0}g_{\alpha 0}
+ \frac{1}{9m^4}k^2q^2g_{\alpha\beta}
- \frac{7}{45m^4}k^2q_{0}^2g_{\alpha\beta}
\]
\[
+  \frac{1}{15m^4}(kq)^2g_{\beta
0}g_{\alpha 0} - \frac{7}{45m^4} (kq)^2g_{\alpha\beta} -
\frac{1}{15m^4}(kq)k_{\beta}q_{0}g_{\alpha 0}
\]
\[
- \frac{1}{15m^4}(kq)k_{\alpha}q_{0}g_{\beta 0} -
\frac{7}{45m^4}k_{\beta}k_{\alpha}q^2 + \frac{1}{15m^4}k_{\beta}k_{\alpha}
q_{0}^2.
\]

It is easy to see that this last expression coincides exactly with the
one corresponding to the Euler-Heisenberg Lagrangian (for the
Euler-Heisenberg expression see, e.g. \cite{akhb}).

\section{CALCULATION OF THE CONTRIBUTION TO THE LAMB SHIFT INDUCED BY THE
LIGHT BY LIGHT SCATTERING INSERTION}

\subsection{LADDER DIAGRAM}

We start the calculation of the contribution to the energy shift induced by the
ladder diagrams in Fig. 1 by multiplication of the respective  part of the
light by light scattering tensor by the numerator of the electron factor in
\eq{ladlght}\footnote{Additional factor 2 is inserted to take into account
both ladder diagrams in Fig.1}. Thus we obtain

\beq
L_l=2A_{\alpha\beta}{\tilde{\cal L}}_{\alpha\beta 00}^{sub}
\eeq
\[
=-2\int_0^1dx\int_0^1du\{\frac{1-u}{\Delta}F^l_1 +\frac{2u(1-u)}{\Delta^2}
F^l_2\},
\]
where

\beq
F^l_1= - 2k^2q_0uvx(1-u) - 6k^2q_0uv(ux)
+ 6k^2q_0u(ux)^2 - 4k^2q_0u(ux)
\eeq
\[
- k^2q_0vx(1-u) + k^2q_0x(1-u)(ux) + 2k^2q_0(1-u)(ux)^2 - 4k^2mu(ux)^2
\]
\[
+ 2k^2mu(ux) + 32(kq)q_0u(ux)[u(1-x)] - 2(kq)q_0u(ux) - 2(kq)q_0u[u(1-x)]
\]
\[
- 2(kq)q_0u + 4(kq)q_0x1(1-u)(ux) + 8(kq)q_0(1-u)(ux)[u(1-x)]
\]
\[
- 8(kq)mu(ux)[u(1-x)] + 2(kq)mu(ux) + 2(kq)mu[u(1-x)] + 2(kq)mu
\]
\[
- 6q^2q_0su[u(1-x)] - q^2q_0sx1(1-u) - 2q^2q_0s(1-u)[u(1-x)]
+ 10q^2q_0u[u(1-x)]^2
\]
\[
- 2q^2q_0u[u(1-x)] + q^2q_0x1(1-u)[u(1-x)] + 2q^2q_0(1-u)[u(1-x)]^2
\]
\[
- 4q^2mu[u(1-x)]^2 + 2q^2mu[u(1-x)] - 8q_0^3u[u(1-x)]^2 - 4q_0^3u[u(1-x)]
\]
\[
+ 8q_0^2mu[u(1-x)]^2 + 4q_0^2mu[u(1-x)],
\]

\medskip

\beq
F^l_2=k^4q_0v^2(ux)^2 - k^4mv^2(ux)^2 + k^4m(ux)^4 - k^4m(ux)^3
\eeq
\[
- 4k^2(kq)q_0v(ux)^2[u(1-x)] - 4k^2(kq)q_0(ux)^3[u(1-x)] + k^2(kq)q_0(ux)^3
\]
\[
+ 3k^2(kq)q_0(ux)^2[u(1-x)] - k^2(kq)q_0(ux)^2 + 4k^2(kq)mv(ux)^2[u(1-x)]
\]
\[
+ 4k^2(kq)m(ux)^3[u(1-x)] - k^2(kq)m(ux)^3 - 3k^2(kq)m(ux)^2[u(1-x)]
\]
\[
+ k^2(kq)m(ux)^2 + 2k^2q^2q_0sv(ux)[u(1-x)] - k^2q^2q_0(ux)^2[u(1-x)]^2
\]
\[
+ k^2q^2q_0(ux)^2[u(1-x)] + 2k^2q^2q_0(ux)[u(1-x)]^2 - 2k^2q^2q_0(ux)[u(1-x)]
\]
\[
- 2k^2q^2msv(ux)[u(1-x)] + 2k^2q^2m(ux)^2[u(1-x)]^2 - k^2q^2m(ux)^2[u(1-x)]
\]
\[
- k^2q^2m(ux)[u(1-x)]^2 + 2k^2q^2m(ux)[u(1-x)] - 2k^2q_0^3(ux)^2[u(1-x)]^2
\]
\[
+ 2k^2q_0^3(ux)^2[u(1-x)] - 2k^2q_0^2m(ux)^2[u(1-x)]^2
- 2k^2q_0^2m(ux)^2[u(1-x)]
\]
\[
- 2k^2q_0m^2v(ux)+ k^2q_0m^2(ux)^2 - 2k^2q_0m^2(ux)
+ 2k^2m^3v(ux) - 2k^2m^3(ux)^2
\]
\[
+ k^2m^3(ux) - 4(kq)^2q_0(ux)^2[u(1-x)]^2 + 2(kq)^2q_0(ux)^2[u(1-x)]
\]
\[
+ 2(kq)^2q_0(ux)[u(1-x)]^2
- 2(kq)^2m(ux)^2[u(1-x)] - 2(kq)^2m(ux)[u(1-x)]^2
\]
\[
- 4(kq)q^2q_0s(ux)[u(1-x)]^2 - 6(kq)q^2q_0(ux)[u(1-x)]^3
\]
\[
+ 3(kq)q^2q_0(ux)[u(1-x)]^2+ (kq)q^2q_0[u(1-x)]^3 - (kq)q^2q_0[u(1-x)]^2
\]
\[
+ 4(kq)q^2ms(ux)[u(1-x)]^2+ 4(kq)q^2m(ux)[u(1-x)]^3
\]
\[
- 3(kq)q^2m(ux)[u(1-x)]^2 - (kq)q^2m[u(1-x)]^3 + (kq)q^2m[u(1-x)]^2
\]
\[
+ 4(kq)q_0^3(ux)[u(1-x)]^3 - 4(kq)q_0^2m(ux)[u(1-x)]^3
+ 10(kq)q_0m^2(ux)[u(1-x)]
\]
\[
- (kq)q_0m^2(ux) - (kq)q_0m^2[u(1-x)] - (kq)q_0m^2 - 8(kq)m^3(ux)[u(1-x)]
\]
\[
+ (kq)m^3(ux) + (kq)m^3[u(1-x)] + (kq)m^3 + q^4q_0s^2[u(1-x)]^2
- q^4q_0[u(1-x)]^4
\]
\[
+ q^4q_0[u(1-x)]^3 - q^4ms^2[u(1-x)]^2 + q^4m[u(1-x)]^4 - q^4m[u(1-x)]^3
\]
\[
+ 2q^2q_0^3[u(1-x)]^4 - 2q^2q_0^3[u(1-x)]^3 - 2q^2q_0^2m[u(1-x)]^4
+ 2q^2q_0^2m[u(1-x)]^3
\]
\[
- 2q^2q_0m^2s[u(1-x)] + 2q^2q_0m^2[u(1-x)]^2 - q^2q_0m^2[u(1-x)]
+ 2q^2m^3s[u(1-x)]
\]
\[
- 2q^2m^3[u(1-x)]^2 + q^2m^3[u(1-x)] - 2q_0^3m^2[u(1-x)]^2
- 2q_0^3m^2[u(1-x)]
\]
\[
+ 2q_0^2m^3[u(1-x)]^2 + 2q_0^2m^3[u(1-x)].
\]

Next we are going to calculate the upper loop momentum integral in Fig.$1a,b$.
Due to two photon propagators this integral contains a factor $q^4$ in the
denominator and one may worry about its convergence at small loop momenta.
We have checked in the previous section that the light by light scattering
tensor supplies at least two powers of loop momentum $q$ in the
low-frequency region making the loop integration completely safe. However,
the integral over the upper loop momentum $q$ will be taken prior to
integration over the Feynman parameters in the light by light scattering
loop and in this case the problem of poor low $q$ behavior  of
the integrand resurrects. In order to make all intermediate integrations
infrared safe we temporarily introduce an intermediate photon mass $\lambda$
for the photons in the upper loop. All calculations both for the ladder and
for the crossed diagrams  will be performed with this nonvanishing photon
mass. The final expression for the energy shift will admit the limit of
vanishing photon mass and will be convergent even in this limit.

Let us combine both electron denominators and photon denominators in the
upper loop in Fig.$1$. To facilitate transformations we introduce new a
notation

\beq  \label{repres}
\Delta=-\frac{1}{\gamma}(q^2+\alpha k^2-2\beta (kq)-\gamma m^2),
\eeq
where

\[
\alpha=\frac{x(1-ux)}{(1-x)[1-u(1-x)]},
\]
\[
\beta=\frac{ux}{1-u(1-x)},
\]
\[
\gamma=\frac{1}{u(1-x)[1-u(1-x)]}.
\]

Then

\beq
(1-t)[(1-y)(q^2-\lambda^2)+y(q^2+2mq_0)]+t(-\gamma\Delta)=q'^2-d_\lambda^2,
\eeq

where

\[
q'=q+y(1-t)m-\beta tk,
\]
\[
d_\lambda^2=-k^2t(\alpha-\beta^2t)+m^2[y^2(1-t)^2+\gamma
t]+\lambda^2(1-t)(1-y).
\]

Hence,

\beq      \label{comb}
\frac{1}{(q^2-\lambda^2)^2(q^2+2mq_0)\Delta}=-6\gamma
\int_0^1dy(1-y)\int_0^1dt(1-t)^2\frac{1}{(q'^2-d_\lambda^2)^4},
\eeq
\[
\frac{1}{(q^2-\lambda^2)^2(q^2+2mq_0)\Delta^2}=24\gamma^2
\int_0^1dy(1-y)\int_0^1dtt(1-t)^2\frac{1}{(q'^2-d_\lambda^2)^5}.
\]

Next euclidean rotation is performed followed by calculation of the integral
over $q_E'$ \footnote{Subscript $E$ is omitted below.} for the contribution
of the ladder diagram to the energy shift in \eq{lamb3}.

\beq          \label{int2}
\Delta E_l=\frac{\alpha^2(Z\alpha)^5}{\pi n^3}m
(\frac{m_r}{m})^3\frac{24}{\pi^2}\alpha^2(Z\alpha)^5m(\frac{m_r}{m})^3
\int\frac{d^3k}{4\pi k^4}\int_0^1dx
\eeq
\[
\int_0^1du\gamma
\int_0^1dy(1-y)\int_0^1dt(1-t)^2\int_0^1dx\int_0^1du\int dq'^2q'^2
\]
\[
\{(1-u)\frac{F^l_1(q=q'-y(1-t)m+\beta tk)}{(q'^2+d_\lambda^2)^4}
\]
\[
+ 8\gamma u(1-u)t\frac{F^l_2(q=q'-y(1-t)m+\beta tk)}{(q'^2+d_\lambda^2)^5}\}.
\]

Explicit expressions for the numerators in the integrand (after Wick
rotation) have the form

\beq
2F^l_1(q')=-q'^2mP^l_1 + 2k^2mP^l_2 + 2m^3[y(1-t)]^2P^l_3,
\eeq

where

\beq
P^l_1=18su[u(1-x)][y(1-t)] + 3s(1-x)(1-u)[y(1-t)]
\eeq
\[
+ 6s(1-u)[u(1-x)][y(1-t)] - 18u[u(1-x)]^2[y(1-t)] - 4u[u(1-x)]^2
\]
\[
+ 12u[u(1-x)][y(1-t)] + 6u[u(1-x)] - 3(1-x)(1-u)[u(1-x)][y(1-t)]
\]
\[
- 6(1-u)[u(1-x)]^2[y(1-t)],
\]

\medskip

\beq
P^l_2=6su[u(1-x)][y(1-t)](\beta t)^2 + s(1-x)(1-u)[y(1-t)](\beta t)^2
\eeq
\[
+ 2s(1-u)[u(1-x)][y(1-t)](\beta t)^2 + 2uvx(1-u)[y(1-t)] + 6uv(ux)[y(1-t)]
\]
\[
- 6u(ux)^2[y(1-t)] - 4u(ux)^2 - 32u(ux)[u(1-x)][y(1-t)](\beta t)
\]
\[
- 8u(ux)[u(1-x)](\beta t)
+ 2u(ux)[y(1-t)](\beta t) + 4u(ux)[y(1-t)] + 2u(ux)(\beta t)
\]
\[
+ 2u(ux) - 10u[u(1-x)]^2[y(1-t)](\beta t)^2 - 4u[u(1-x)]^2(\beta t)^2
\]
\[
+ 2u[u(1-x)][y(1-t)](\beta t)^2
+ 2u[u(1-x)][y(1-t)](\beta t) + 2u[u(1-x)](\beta t)^2
\]
\[
+ 2u[u(1-x)](\beta t) + 2u[y(1-t)](\beta t) + 2u(\beta t) + vx(1-u)[y(1-t)]
\]
\[
- x(1-u)(ux)[y(1-t)] - 4(1-x)(1-u)(ux)[y(1-t)](\beta t)
\]
\[
- (1-x)(1-u)[u(1-x)][y(1-t)](\beta t)^2 - 2(1-u)(ux)^2[y(1-t)]
\]
\[
- 8(1-u)(ux)[u(1-x)][y(1-t)](\beta t) - 2(1-u)[u(1-x)]^2[y(1-t)](\beta t)^2,
\]

\medskip

\beq
P^l_3=6su[u(1-x)][y(1-t)] + s(1-x)(1-u)[y(1-t)]
\eeq
\[
+ 2s(1-u)[u(1-x)][y(1-t)] - 2u[u(1-x)]^2[y(1-t)] + 4u[u(1-x)]^2
\]
\[
+ 6u[u(1-x)][y(1-t)] + 6u[u(1-x)] - (1-x)(1-u)[u(1-x)][y(1-t)]
\]
\[
 - 2(1-u)[u(1-x)]^2[y(1-t)].
\]

The second term in the integrand in \eq{int2} has the  following form

\beq
2F^l_2=\frac{1}{3}q'^4m[u(1-x)]^2T^l_4 - q'^2k^2m[u(1-x)]T^l_1
- q'^2m^3[u(1-x)]T^l_5
\eeq
\[
+ 2k^4mT^l_2+ 2k^2m^3T^l_3 + 2m^5[u(1-x)][y(1-t)]^2T^l_6,
\]
where

\beq
T^l_4 = - 12s^2[y(1-t)] - 6s^2 + 2[u(1-x)]^2[y(1-t)]
\eeq
\[
+ 3[u(1-x)]^2 - 2[u(1-x)][y(1-t)] - 3[u(1-x)],
\]

\medskip

\beq
T^l_1 =  - 8s^2[u(1-x)][y(1-t)](\beta t)^2 - 6s^2[u(1-x)](\beta t)^2
\eeq
\[
- 6sv(ux)[y(1-t)] - 4sv(ux) + 16s(ux)[u(1-x)][y(1-t)](\beta t)
\]
\[
+ 12s(ux)[u(1-x)](\beta t) + 8(ux)^2[u(1-x)][y(1-t)] + 3(ux)^2[u(1-x)]
\]
\[
- 7(ux)^2[y(1-t)] - 4(ux)^2 + 18(ux)[u(1-x)]^2[y(1-t)](\beta t)
\]
\[
+ 10(ux)[u(1-x)]^2(\beta t) - 12(ux)[u(1-x)][y(1-t)](\beta t)
- 7(ux)[u(1-x)][y(1-t)]
\]
\[
- 9(ux)[u(1-x)](\beta t) - 3(ux)[u(1-x)] + 6(ux)[y(1-t)] + 4(ux)
\]
\[
+ 5[u(1-x)]^3[y(1-t)](\beta t)^2 + 5[u(1-x)]^3(\beta t)^2
- 5[u(1-x)]^2[y(1-t)](\beta t)^2
\]
\[
- 4[u(1-x)]^2[y(1-t)](\beta t)
- 5[u(1-x)]^2(\beta t)^2 - 3[u(1-x)]^2(\beta t)
\]
\[
+ 4[u(1-x)][y(1-t)](\beta t) + 3[u(1-x)](\beta t),
\]

\medskip

\beq
T^l_5 = - 8s^2[u(1-x)][y(1-t)]^3 - 6s^2[u(1-x)][y(1-t)]^2
\eeq
\[
+ 6s[y(1-t)] + 4s - 5[u(1-x)]^3[y(1-t)]^3 - 3[u(1-x)]^3[y(1-t)]^2
\]
\[
+ 5[u(1-x)]^2[y(1-t)]^3
+ 3[u(1-x)]^2[y(1-t)]^2 - 3[u(1-x)][y(1-t)]
\]
\[
- 3[u(1-x)] + 6[y(1-t)] + 3,
\]

\medskip

\beq
T^l_2 =- s^2[u(1-x)]^2[y(1-t)](\beta t)^4 - s^2[u(1-x)]^2(\beta t)^4
\eeq
\[
- 2sv(ux)[u(1-x)][y(1-t)](\beta t)^2 - 2sv(ux)[u(1-x)](\beta t)^2
\]
\[
+ 4s(ux)[u(1-x)]^2[y(1-t)](\beta t)^3
+ 4s(ux)[u(1-x)]^2(\beta t)^3
\]
\[
- v^2(ux)^2[y(1-t)] - v^2(ux)^2 + 4v(ux)^2[u(1-x)][y(1-t)](\beta t)
\]
\[
+ 4v(ux)^2[u(1-x)](\beta t) + (ux)^4 + 4(ux)^3[u(1-x)][y(1-t)](\beta t)
\]
\[
 + 4(ux)^3[u(1-x)](\beta t) - (ux)^3[y(1-t)](\beta t) - (ux)^3(\beta t)
- (ux)^3
\]
\[
+ 5(ux)^2[u(1-x)]^2[y(1-t)](\beta t)^2
+ 2(ux)^2[u(1-x)]^2(\beta t)^2
\]
\[
- 3(ux)^2[u(1-x)][y(1-t)](\beta t)^2 - 3(ux)^2[u(1-x)][y(1-t)](\beta t)
\]
\[
- 3(ux)^2[u(1-x)](\beta t)^2 - 3(ux)^2[u(1-x)](\beta t)
+ (ux)^2[y(1-t)](\beta t)
\]
\[
+ (ux)^2(\beta t) + 6(ux)[u(1-x)]^3[y(1-t)](\beta t)^3
+ 4(ux)[u(1-x)]^3(\beta t)^3
\]
\[
- 3(ux)[u(1-x)]^2[y(1-t)](\beta t)^3 - 4(ux)[u(1-x)]^2[y(1-t)](\beta t)^2
\]
\[
- 3(ux)[u(1-x)]^2(\beta t)^3 - 3(ux)[u(1-x)]^2(\beta t)^2
+ 2(ux)[u(1-x)][y(1-t)](\beta t)^2
\]
\[
+ 2(ux)[u(1-x)](\beta t)^2 + [u(1-x)]^4[y(1-t)](\beta t)^4
+ [u(1-x)]^4(\beta t)^4
\]
\[
- [u(1-x)]^3[y(1-t)](\beta t)^4
- [u(1-x)]^3[y(1-t)](\beta t)^3 - [u(1-x)]^3(\beta t)^4
\]
\[
- [u(1-x)]^3(\beta t)^3
+ [u(1-x)]^2[y(1-t)](\beta t)^3 + [u(1-x)]^2(\beta t)^3,
\]

\medskip

\beq
T^l_3= - 2s^2[u(1-x)]^2[y(1-t)]^3(\beta t)^2
- 2s^2[u(1-x)]^2[y(1-t)]^2(\beta t)^2
\eeq
\[
- 2sv(ux)[u(1-x)][y(1-t)]^3 - 2sv(ux)[u(1-x)][y(1-t)]^2
\]
\[
+ 4s(ux)[u(1-x)]^2[y(1-t)]^3(\beta t) + 4s(ux)[u(1-x)]^2[y(1-t)]^2(\beta t)
\]
\[
+ 2s[u(1-x)][y(1-t)](\beta t)^2 + 2s[u(1-x)](\beta t)^2
+ 2v(ux)[y(1-t)] + 2v(ux)
\]
\[
+ 3(ux)^2[u(1-x)]^2[y(1-t)]^3 - 3(ux)^2[u(1-x)][y(1-t)]^3
\]
\[
- 3(ux)^2[u(1-x)][y(1-t)]^2 - (ux)^2[y(1-t)] - 2(ux)^2
\]
\[
+ 2(ux)[u(1-x)]^3[y(1-t)]^3(\beta t)
- 3(ux)[u(1-x)]^2[y(1-t)]^3(\beta t)
\]
\[
- 2(ux)[u(1-x)]^2[y(1-t)]^3 - 3(ux)[u(1-x)]^2[y(1-t)]^2(\beta t)
\]
\[
- (ux)[u(1-x)]^2[y(1-t)]^2 + 2(ux)[u(1-x)][y(1-t)]^3
+ 2(ux)[u(1-x)][y(1-t)]^2
\]
\[
- 10(ux)[u(1-x)][y(1-t)](\beta t) - 8(ux)[u(1-x)](\beta t)
+ (ux)[y(1-t)](\beta t)
\]
\[
 + 2(ux)[y(1-t)] + (ux)(\beta t) + (ux) - [u(1-x)]^3[y(1-t)]^3(\beta t)
\]
\[
- [u(1-x)]^3[y(1-t)]^2(\beta t)
+ [u(1-x)]^2[y(1-t)]^3(\beta t) + [u(1-x)]^2[y(1-t)]^2(\beta t)
\]
\[
- 2[u(1-x)]^2[y(1-t)](\beta t)^2 - 2[u(1-x)]^2(\beta t)^2
+ [u(1-x)][y(1-t)](\beta t)^2
\]
\[
+ [u(1-x)][y(1-t)](\beta t)
+ [u(1-x)](\beta t)^2 + [u(1-x)](\beta t) + [y(1-t)](\beta t) + (\beta t),
\]

\medskip

\beq
T^l_6 = - s^2[u(1-x)][y(1-t)]^3 - s^2[u(1-x)][y(1-t)]^2
\eeq
\[
+ 2s[y(1-t)] + 2s - [u(1-x)]^3[y(1-t)]^3 - [u(1-x)]^3[y(1-t)]^2
\]
\[
+ [u(1-x)]^2[y(1-t)]^3 + [u(1-x)]^2[y(1-t)]^2 + 3[y(1-t)] + 3.
\]

After integration we obtain

\beq            \label{infrared}
\Delta E_{l}=\frac{\alpha^2(Z\alpha)^5}{\pi n^3}m
(\frac{m_r}{m})^3\frac{12}{\pi^2}
\int_0^1dx\int_0^1du(1-u)\gamma
\eeq
\[
\int_0^1dy(1-y)\int_0^1dt(1-t)^2\int\frac{d|\bf k|}{{\bf k}^2}
\{4\gamma t{\bf k}^4\frac{muT^l_2}{3d_\lambda^6}
\]
\[
-{\bf k}^2[\frac{mP^l_2-2\gamma tm[u(1-x)]uT^l_1}{3d^4}+4\gamma t
\frac{m^3uT^l_3}{3d_\lambda^6}]
\]
\[
-\frac{mP^l_1}{3d_\lambda^2}+\frac{m^3[y(1-t)]^2P^l_3}{3d_\lambda^4}
+ \frac{4}{3}\gamma t\frac{m[u(1-x)]^2uT^l_4}{2d_\lambda^2}
\]
\[
-4\gamma t\frac{m^3[u(1-x)]uT^l_5}{6d_\lambda^4}
+4\gamma t\frac{m^5[u(1-x)]u[y(1-t)]^2T^l_6}{3d_\lambda^6} \}.
\]

Integration over $|{\bf k}|$ may easily be performed since it is
essentially one-dimensional due to spacelike nature of vector $k$.

It is convenient to introduce the following notation

\beq
d_\lambda^2={\bf k}^2t(\alpha-\beta^2t)+m^2[y^2(1-t)^2+\gamma t]+\lambda^2
(1-t)(1-y)
\eeq
\[
\equiv\rho \{{\bf k}^2+\frac{m^2[y^2(1-t)^2+\gamma t]+\lambda^2(1-t)(1-y)}
{\rho}\}
\]
\[
=\rho ({\bf k}^2+\omega_\lambda^2),
\]
where

\beq
\rho\equiv t(\alpha-\beta^2t),
\eeq
\[
\omega_\lambda^2\equiv \frac{m^2[y^2(1-t)^2
+\gamma t]+\lambda^2(1-t)(1-y)}{\rho}.
\]

Integration over ${\bf |k|}$ in \eq{infrared} is impeded
by the apparent infrared divergence, which is connected with the apparent
constant  asymptote  of the electron factor. \footnote{We call here electron
factor the product of the electron factor defined above and
the light by light scattering tensor. Normalization of this electron factor
differs from the one used in our previous works \cite{eg3,eg4} on the
contributions to the Lamb shift; in terms of the old normalization the
asymptote
we are discussing now is proportional to ${\bf k}^2$.} However, one expects
that the leading low frequency term in this asymptote is proportional to
${\bf k}^2$ (if this is not the case the graphs under consideration would
produce contributions of even lower order in $Z\alpha$ which is well known to
be the wrong conclusion). Surely, only the total electron-line factor should be
proportional ${\bf k}^2$ in the low frequency limit and not the part taken
into account in \eq{infrared}. These considerations give another
opportunity to
check the validity of all transformations above. One has to calculate all parts
of the electron-line factor and then to calculate its low frequency asymptote.
If it turns out that the electron-line factor vanishes as ${\bf k}^2$
when ${\bf k}$ is small then there are really no difficulties in the treatment
 of integrals of the sort contained in \eq{infrared}; one has simply to perform
subtraction of the leading low frequency terms in the electron-line factor
since these terms cancel in any case in the total electron-line factor. We
will check the disappearance of the apparent leading term in the asymptote
of the total electron factor in the next Section.

Subtraction of the low frequency part is necessary in the last five terms
of the integrand in \eq{infrared}. The numerators of these terms are
independent of momentum $k$ and subtraction may be easily performed
with the help of the identities

\beq
\frac{1}{d_\lambda^2({\bf k})}-\frac{1}{d_\lambda^2(0)}
=-\frac{{\bf k}^2}{\omega_\lambda^2d_\lambda^2({\bf
k})}
\equiv -\frac{{\bf k}^2}{\rho\omega_\lambda^2({\bf k}^2+\omega_\lambda^2)},
\eeq
\[
\frac{1}{d_\lambda^4({\bf k})}-\frac{1}{d_\lambda^4(0)}=-\frac{{\bf
k}^2}{\rho\omega_\lambda^4d_\lambda^2({\bf k})}-\frac{{\bf
k}^2}{\omega_\lambda^2d_\lambda^4({\bf k})}
\]
\[
\equiv -\frac{{\bf k}^2}{\rho^2\omega_\lambda^4({\bf
k}^2+\omega_\lambda^2)}-\frac{{\bf k}^2}{\rho^2\omega_\lambda^2({\bf
k}^2+\omega_\lambda^2)^2},
\]
\[
\frac{1}{d_\lambda^6({\bf k})}-\frac{1}{d_\lambda^6(0)}=-\frac{{\bf
k}^2}{\rho^2\omega_\lambda^6d_\lambda^2({\bf k})}-\frac{{\bf
k}^2}{\rho\omega_\lambda^4d_\lambda^4({\bf k})}-\frac{{\bf
k}^2}{\omega_\lambda^2d_\lambda^6({\bf k})}
\]
\[
\equiv -\frac{{\bf k}^2}{\rho^3\omega_\lambda^6({\bf
k}^2+\omega_\lambda^2)}-\frac{{\bf k}^2}{\rho^3\omega_\lambda^4({\bf
k}^2+\omega_\lambda^2)^2}-\frac{{\bf k}^2}{\rho^3\omega_\lambda^2({\bf
k}^2+\omega_\lambda^2)^3}.
\]

Substituting these expressions in  \eq{infrared} we obtain

\beq
\Delta E_{l}^{sub}
=\frac{\alpha^2(Z\alpha)^5}{\pi n^3}m
(\frac{m_r}{m})^3\frac{12}{\pi^2}
\int_0^1dx
\eeq
\[
\int_0^1du(1-u)\gamma
\int_0^1dy(1-y)\int_0^1dt(1-t)^2\int_0^\infty d|\bf k|
\]
\[
\{\frac{4\gamma t}{3\rho^3({\bf k}^2
+\omega_\lambda^2)^3}[-m\omega_\lambda^2uT^l_2
-m^3uT^l_3-\frac{m^5[u(1-x)]u[y(1-t)]^2T^l_6}{\omega_\lambda^2}]
\]
\[
+\frac{1}{3\rho^2({\bf k}^2+\omega_\lambda^2)^2}[
4\gamma t\frac{muT^l_2}{\rho}
-mP^l_2+2\gamma tm[u(1-x)]uT^l_1
-\frac{m^3[y(1-t)]^2P^l_3}{\omega_\lambda^2}
\]
\[
+2\gamma t\frac{m^3[u(1-x)]uT^l_5}{\omega_\lambda^2}
-4\gamma t\frac{m^5[u(1-x)]u[y(1-t)]^2T^l_6}{\rho\omega_\lambda^4}]
\]
\[
+\frac{m}{3\rho\omega^2({\bf k}^2+\omega_\lambda^2)}
[P^l_1
-\frac{m^2[y(1-t)]^2P^l_3}{\rho\omega_\lambda^2}
- 2\gamma t[u(1-x)]^2uT^l_4
\]
\[
+2\gamma t\frac{m^2[u(1-x)]uT^l_5}{\rho\omega_\lambda^2}
-4\gamma
t\frac{m^4[u(1-x)]u[y(1-t)]^2T^l_6}{\rho^2\omega_\lambda^4}] \}.
\]

Integrating next over $|{\bf k}|$ we  obtain

\beq     \label{laddercontr}
\Delta E_{l}^{sub}=\frac{\alpha^2(Z\alpha)^5}{\pi n^3}m
(\frac{m_r}{m})^3\frac{12}{\pi}
\int_0^1dx
\eeq
\[
\int_0^1du(1-u)\gamma
\int_0^1dy(1-y)\int_0^1dt(1-t)^2
\]
\[
\{\frac{\gamma t}{4\rho^3\omega_\lambda^5}[-m\omega_\lambda^2uT^l_2
-m^3uT^l_3-\frac{m^5[u(1-x)]u[y(1-t)]^2T^l_6}{\omega_\lambda^2}]
\]
\[
+\frac{1}{12\rho^2\omega_\lambda^3}[
4\gamma t\frac{muT^l_2}{\rho}
-mP^l_2+2\gamma tm[u(1-x)]uT^l_1
-\frac{m^3[y(1-t)]^2P^l_3}{\omega_\lambda^2}
\]
\[
+2\gamma t\frac{m^3[u(1-x)]uT^l_5}{\omega_\lambda^2}
-4\gamma t\frac{m^5[u(1-x)]u[y(1-t)]^2T^l_6}{\rho\omega_\lambda^4}]
\]
\[
+\frac{m}{6\rho\omega_\lambda^3}
[P^l_1
-\frac{m^2[y(1-t)]^2P^l_3}{\rho\omega_\lambda^2}
- 2\gamma t[u(1-x)]^2uT^l_4
\]
\[
+2\gamma t\frac{m^2[u(1-x)]uT^l_5}{\rho\omega_\lambda^2}
-4\gamma
t\frac{m^4[u(1-x)]u[y(1-t)]^2T^l_6}{\rho^2\omega_\lambda^4}] \}.
\]

\subsection{CROSSED DIAGRAM}

Consideration of the crossed diagram contribution to the Lamb shift follows the
same lines as in the case of the ladder diagram. First we multiply the
respective
part of the light by light scattering tensor by the numerator of the
electron factor in \eq{ladlght} and obtain

\beq
L_c=A_{\alpha\beta}{\tilde{\cal C}}_{\alpha\beta 00}^{sub}
\eeq
\[
=-\int_0^1dx\int_0^1du\int_0^1dz
\{\frac{u}{\Delta_z}2F^c_1 +\frac{uz^2}{\Delta^2_z}2F^c_2\},
\]

where

\beq
F^c_1= - 8k^2q_0a^2z^2 + 8k^2q_0az^2 + 6k^2q_0v^2z^3(1-z)
\eeq
\[
+ 4k^2q_0vz^3(ux) + 2k^2q_0v(ux) + 2k^2q_0z^4(ux)^2 - 4k^2q_0z^4(ux)
+ 2k^2q_0z^4
\]
\[
- 2k^2q_0z^3(1-z)(ux)^2 + 4k^2q_0z^3(1-z)(ux)
- 2k^2q_0z^3(1-z) - k^2q_0z^2
\]
\[
- 2k^2q_0z(ux)^2 + 4k^2q_0z(ux) - 2k^2q_0z + 2k^2q_0(1-z)(ux)^2
- 4k^2q_0(1-z)(ux)
\]
\[
+ 2k^2q_0(1-z) - 2k^2a^2mz^2
+ 2k^2amz^2 - k^2mz^2 - 12(kq)q_0abz^2 + 6(kq)q_0az^2
\]
\[
+ 6(kq)q_0bz^2 + 12(kq)q_0svz^3(1-z) + 4(kq)q_0z^4(ux)[u(1-x)]
- 4(kq)q_0z^4(ux)
\]
\[
- 4(kq)q_0z^4[u(1-x)] + 4(kq)q_0z^4 - 4(kq)q_0z^3(1-z)(ux)[u(1-x)]
\]
\[
+ 4(kq)q_0z^3(1-z)(ux)
+ 4(kq)q_0z^3(1-z)[u(1-x)] - 4(kq)q_0z^3(1-z)
\]
\[
- 8(kq)q_0z^3(ux)[u(1-x)] - 3(kq)q_0z^2 - 4(kq)q_0z(ux)[u(1-x)]
+ 4(kq)q_0z(ux)
\]
\[
+ 4(kq)q_0z[u(1-x)] - 4(kq)q_0z
+ 4(kq)q_0(1-z)(ux)[u(1-x)]
\]
\[
- 4(kq)q_0(1-z)(ux) - 4(kq)q_0(1-z)[u(1-x)] + 4(kq)q_0(1-z)
\]
\[
- 4(kq)q_0(ux)[u(1-x)] - 4(kq)abmz^2 + 2(kq)amz^2
+ 2(kq)bmz^2 - (kq)mz^2
\]
\[
- 10q^2q_0b^2z^2 + 10q^2q_0bz^2 + 6q^2q_0s^2z^3(1-z) + 4q^2q_0sz^3[u(1-x)]
\]
\[
+ 2q^2q_0s[u(1-x)] + 2q^2q_0z^4[u(1-x)]^2 - 4q^2q_0z^4[u(1-x)] + 2q^2q_0z^4
\]
\[
- 2q^2q_0z^3(1-z)[u(1-x)]^2 + 4q^2q_0z^3(1-z)[u(1-x)]
- 2q^2q_0z^3(1-z) - 2q^2q_0z^2
\]
\[
- 2q^2q_0z[u(1-x)]^2 + 4q^2q_0z[u(1-x)] - 2q^2q_0z + 2q^2q_0(1-z)[u(1-x)]^2
\]
\[
- 4q^2q_0(1-z)[u(1-x)] + 2q^2q_0(1-z) - 2q^2b^2mz^2 + 2q^2bmz^2
- q^2mz^2 + 4q_0^3b^2z^2
\]
\[
- 4q_0^3bz^2 + 2q_0^3z^2
- 4q_0^2b^2mz^2 + 4q_0^2bmz^2 - 2q_0^2mz^2,
\]

\medskip

\beq
F^c_2=2k^4q_0a^4 - 4k^4q_0a^3 + 2k^4q_0a^2 - 2k^4q_0v^4z^2(1-z)^2
\eeq
\[
- 4k^4q_0v^3z^2(1-z)(ux) - 2k^4q_0v^2z^2(ux)^2
+ k^4a^4m - 2k^4a^3m + k^4a^2m
\]
\[
- k^4mv^4z^2(1-z)^2 - 2k^4mv^3z^2(1-z)(ux) - k^4mv^2z^2(ux)^2
+ 4k^2(kq)q_0a^3b
\]
\[
- 2k^2(kq)q_0a^3 - 6k^2(kq)q_0a^2b
+ 3k^2(kq)q_0a^2 + 2k^2(kq)q_0ab - k^2(kq)q_0a
\]
\[
- 8k^2(kq)q_0sv^3z^2(1-z)^2 - 8k^2(kq)q_0sv^2z^2(1-z)(ux)
\]
\[
+ 8k^2(kq)q_0v^2z^2(1-z)(ux)[u(1-x)] + 8k^2(kq)q_0vz^2(ux)^2[u(1-x)]
\]
\[
+ 4k^2(kq)a^3bm - 2k^2(kq)a^3m - 6k^2(kq)a^2bm + 3k^2(kq)a^2m + 2k^2(kq)abm
\]
\[
- k^2(kq)am - 4k^2(kq)msv^3z^2(1-z)^2 - 4k^2(kq)msv^2z^2(1-z)(ux)
\]
\[
+ 4k^2(kq)mv^2z^2(1-z)(ux)[u(1-x)] + 4k^2(kq)mvz^2(ux)^2[u(1-x)]
\]
\[
+ 4k^2q^2q_0a^2b^2 - 4k^2q^2q_0a^2b + 2k^2q^2q_0a^2
- 4k^2q^2q_0ab^2 + 4k^2q^2q_0ab - 2k^2q^2q_0a
\]
\[
+ k^2q^2q_0b^2 - k^2q^2q_0b - 4k^2q^2q_0s^2v^2z^2(1-z)^2
- 4k^2q^2q_0s^2vz^2(1-z)(ux)
\]
\[
- 4k^2q^2q_0sv^2z^2(1-z)[u(1-x)]
- 4k^2q^2q_0svz^2(ux)[u(1-x)] + 2k^2q^2a^2b^2m
\]
\[
- 2k^2q^2a^2bm + k^2q^2a^2m - 2k^2q^2ab^2m + 2k^2q^2abm - k^2q^2am
+ k^2q^2b^2m - k^2q^2bm
\]
\[
- 2k^2q^2ms^2v^2z^2(1-z)^2 - 2k^2q^2ms^2vz^2(1-z)(ux)
- 2k^2q^2msv^2z^2(1-z)[u(1-x)]
\]
\[
- 2k^2q^2msvz^2(ux)[u(1-x)] - 4k^2q_0^3a^2b^2
+ 4k^2q_0^3a^2b - 2k^2q_0^3a^2 + 4k^2q_0^3ab^2
\]
\[
- 4k^2q_0^3ab + 2k^2q_0^3a + 2k^2q_0^2a^2m - 2k^2q_0^2am
- 4k^2q_0a^2m^2 + 4k^2q_0am^2
\]
\[
+ 2k^2q_0m^2v^2z^2(1-z) + 2k^2q_0m^2v^2z(1-z)^2 + 2k^2q_0m^2v^2z(1-z)
\]
\[
+ 2k^2q_0m^2vz^2(ux) + 2k^2q_0m^2vz(1-z)(ux)
+ 2k^2q_0m^2vz(ux) - k^2q_0m^2
\]
\[
- 2k^2a^2m^3  + 2k^2am^3
+ k^2m^3v^2z^2(1-z) + k^2m^3v^2z(1-z)^2 + k^2m^3v^2z(1-z)
\]
\[
+ k^2m^3vz^2(ux) + k^2m^3vz(1-z)(ux) + k^2m^3vz(ux)
- k^2m^3
\]
\[
- 8(kq)^2q_0s^2v^2z^2(1-z)^2 + 16(kq)^2q_0svz^2(1-z)(ux)[u(1-x)]
\]
\[
- 8(kq)^2q_0z^2(ux)^2[u(1-x)]^2
+ 4(kq)^2a^2b^2m - 4(kq)^2a^2bm - 4(kq)^2ab^2m
\]
\[
+ 4(kq)^2abm - 4(kq)^2ms^2v^2z^2(1-z)^2 + 8(kq)^2msvz^2(1-z)(ux)[u(1-x)]
\]
\[
- 4(kq)^2mz^2(ux)^2[u(1-x)]^2 + 4(kq)q^2q_0ab^3 - 6(kq)q^2q_0ab^2
+ 2(kq)q^2q_0ab
\]
\[
- 2(kq)q^2q_0b^3 + 3(kq)q^2q_0b^2
- (kq)q^2q_0b - 8(kq)q^2q_0s^3vz^2(1-z)^2
\]
\[
- 8(kq)q^2q_0s^2vz^2(1-z)[u(1-x)]
+ 8(kq)q^2q_0s^2z^2(1-z)(ux)[u(1-x)]
\]
\[
+ 8(kq)q^2q_0sz^2(ux)[u(1-x)]^2 + 4(kq)q^2ab^3m - 6(kq)q^2ab^2m + 2(kq)q^2abm
\]
\[
- 2(kq)q^2b^3m + 3(kq)q^2b^2m - (kq)q^2bm - 4(kq)q^2ms^3vz^2(1-z)^2
\]
\[
- 4(kq)q^2ms^2vz^2(1-z)[u(1-x)] + 4(kq)q^2ms^2z^2(1-z)(ux)[u(1-x)]
\]
\[
+ 4(kq)q^2msz^2(ux)[u(1-x)]^2 - 4(kq)q_0abm^2 + 2(kq)q_0am^2 + 2(kq)q_0bm^2
\]
\[
+ 4(kq)q_0m^2svz^2(1-z) + 4(kq)q_0m^2svz(1-z)^2 + 4(kq)q_0m^2svz(1-z)
\]
\[
- 4(kq)q_0m^2z^2(ux)[u(1-x)] - 4(kq)q_0m^2z(1-z)(ux)[u(1-x)]
\]
\[
- 4(kq)q_0m^2z(ux)[u(1-x)] - (kq)q_0m^2 - 4(kq)abm^3 + 2(kq)am^3 + 2(kq)bm^3
\]
\[
+ 2(kq)m^3svz^2(1-z) + 2(kq)m^3svz(1-z)^2 + 2(kq)m^3svz(1-z)
\]
\[
- 2(kq)m^3z^2(ux)[u(1-x)] - 2(kq)m^3z(1-z)(ux)[u(1-x)]
\]
\[
- 2(kq)m^3z(ux)[u(1-x)] - (kq)m^3 + 2q^4q_0b^4 - 4q^4q_0b^3 + 2q^4q_0b^2
\]
\[
- 2q^4q_0s^4z^2(1-z)^2- 4q^4q_0s^3z^2(1-z)[u(1-x)]
- 2q^4q_0s^2z^2[u(1-x)]^2 + q^4b^4m
\]
\[
- 2q^4b^3m + q^4b^2m - q^4ms^4z^2(1-z)^2 - 2q^4ms^3z^2(1-z)[u(1-x)]
\]
\[
- q^4ms^2z^2[u(1-x)]^2
- 4q^2q_0b^2m^2 + 4q^2q_0bm^2 + 2q^2q_0m^2s^2z^2(1-z)
\]
\[
+ 2q^2q_0m^2s^2z(1-z)^2 + 2q^2q_0m^2s^2z(1-z) + 2q^2q_0m^2sz^2[u(1-x)]
\]
\[
+ 2q^2q_0m^2sz(1-z)[u(1-x)]
+ 2q^2q_0m^2sz[u(1-x)] - 2q^2q_0m^2 - 2q^2b^2m^3 + 2q^2bm^3
\]
\[
+ q^2m^3s^2z^2(1-z) + q^2m^3s^2z(1-z)^2 + q^2m^3s^2z(1-z)
+ q^2m^3sz^2[u(1-x)]
\]
\[
+ q^2m^3sz(1-z)[u(1-x)]
+ q^2m^3sz[u(1-x)] - q^2m^3 + 2q_0^3m^2 - 2q_0^2m^3.
\]

Next we go to calculation of the upper photon loop in Fig.1c. To facilitate
transformations we introduce new notation  which resembles the one used in
the previous section

\beq  \label{represz}
\Delta_z\equiv m^2-z\{k^2(1-ux)[1-z(1-ux)]+q^2[1-u(1-x)][1-z(1-u(1-x))]
\eeq
\[
+2kq[1-u-z(1-ux)(1-u(1-x))]\}
\]
\[
=-\frac{1}{\gamma_z}[q^2+\alpha_z k^2-2\beta_z (kq)-\gamma_z m^2],
\]

where

\[
\alpha_z=\frac{(1-ux)[1-z(1-ux)]}{[1-u(1-x)][1-z(1-u(1-x))]},
\]
\[
\beta_z=-\frac{[1-u-z(1-ux)(1-u(1-x))]}{[1-u(1-x)][1-z(1-u(1-x))]},
\]
\[
\gamma_z=\frac{1}{z[1-u(1-x)][1-z(1-u(1-x))]}.
\]

Note that

\beq
\alpha_z(z=1)=\alpha,
\eeq
\[
\beta_z(z=1)=\beta,
\]
\[
\gamma_z(z=1)=\gamma.
\]

As was discussed in detail in the case of the ladder diagram we have to
introduce small photon mass $\lambda$ for the photons in the upper
loop to make all intermediate integrations infrared safe.

Then

\beq
(1-t)[(1-y)(q^2-\lambda^2)+y(q^2+2mq_0)]+t(-\gamma_z\Delta_z)
=q''^2-d_{\lambda z}^2,
\eeq

where

\[
q''=q+y(1-t)m-\beta_z tk,
\]
\[
d_{\lambda z}^2=-k^2t(\alpha_z-\beta_z^2t)+m^2[y^2(1-t)^2+\gamma_z t]
+\lambda^2(1-t)(1-y) .
\]

Hence,

\beq      \label{combz}
\frac{1}{(q^2-\lambda^2)^2(q^2+2mq_0)\Delta_z}=-6\gamma_z
\int_0^1dy(1-y)\int_0^1dt(1-t)^2\frac{1}{(q''^2-d_{\lambda z}^2)^4},
\eeq
\[
\frac{1}{(q^2-\lambda^2)^2(q^2+2mq_0)\Delta_z^2}=24\gamma_z^2
\int_0^1dy(1-y)\int_0^1dtt(1-t)^2\frac{1}{(q''^2-d_{\lambda z}^2)^5}.
\]

Next we perform euclidean rotation and calculate the integral over $q_E''$
\footnote{Subscript $E$ is omitted below.} for the contribution of the
crossed diagram to the energy shift in \eq{lamb3}.

\beq          \label{int2z}
\Delta E_c=\frac{\alpha^2(Z\alpha)^5}{\pi n^3}m
(\frac{m_r}{m})^3\frac{12}{\pi^2}
\int\frac{d^3k}{4\pi k^4}\int_0^1dx
\eeq
\[
\int_0^1duu\int_0^1dz\gamma_z
\int_0^1dy(1-y)\int_0^1dt(1-t)^2\int dq''^2q''^2
\]
\[
\{\frac{2F^c_1(q=q''-y(1-t)m+\beta_z tk)}{(q''^2+d_{\lambda z}^2)^4}
+4\gamma_z t\frac{2z^2F^c_2(q=q''-y(1-t)m+\beta_z tk)}{(q''^2
+d_{\lambda z}^2)^5}\}.
\]

Explicit expressions for the numerators (after Wick rotation) in the
integrand have the form

\beq
2F^c_1= -3q''^2mP^c_1 + 2k^2mP^c_2 + 2m^3[y(1-t)]^2P^c_3,
\eeq

where

\beq
P^c_1=8b^2z^2[y(1-t)] - 2b^2z^2 - 8bz^2[y(1-t)] + 2bz^2
\eeq
\[
- 6s^2z^3(1-z)[y(1-t)] - 4sz^3[u(1-x)][y(1-t)] - 2s[u(1-x)][y(1-t)]
\]
\[
- 2z^4[u(1-x)]^2[y(1-t)]
+ 4z^4[u(1-x)][y(1-t)] - 2z^4[y(1-t)]
\]
\[
+ 2z^3(1-z)[u(1-x)]^2[y(1-t)] - 4z^3(1-z)[u(1-x)][y(1-t)]
\]
\[
+ 2z^3(1-z)[y(1-t)] + z^2[y(1-t)] - z^2 + 2z[u(1-x)]^2[y(1-t)]
\]
\[
- 4z[u(1-x)][y(1-t)]
+ 2z[y(1-t)] - 2(1-z)[u(1-x)]^2[y(1-t)]
\]
\[
+ 4(1-z)[u(1-x)][y(1-t)] - 2(1-z)[y(1-t)],
\]

\medskip

\beq
P^c_2=8a^2z^2[y(1-t)] - 2a^2z^2 + 12abz^2[y(1-t)](\beta_z t)
\eeq
\[
- 4abz^2(\beta_z t) - 6az^2[y(1-t)](\beta_z t) - 8az^2[y(1-t)]
+ 2az^2(\beta_z t) + 2az^2
\]
\[
+ 10b^2z^2[y(1-t)](\beta_z t)^2 - 2b^2z^2(\beta_z t)^2
- 10bz^2[y(1-t)](\beta_z t)^2 - 6bz^2[y(1-t)](\beta_z t)
\]
\[
+ 2bz^2(\beta_z t)^2 + 2bz^2(\beta_z t) - 6s^2z^3(1-z)[y(1-t)](\beta_z t)^2
\]
\[
- 12svz^3(1-z)[y(1-t)](\beta_z t) - 4sz^3[u(1-x)][y(1-t)](\beta_z t)^2
\]
\[
- 2s[u(1-x)][y(1-t)](\beta_z t)^2 - 6v^2z^3(1-z)[y(1-t)]
- 4vz^3(ux)[y(1-t)]
\]
\[
- 2v(ux)[y(1-t)] - 2z^4(ux)^2[y(1-t)] - 4z^4(ux)[u(1-x)][y(1-t)](\beta_z t)
\]
\[
+ 4z^4(ux)[y(1-t)](\beta_z t) + 4z^4(ux)[y(1-t)]
- 2z^4[u(1-x)]^2[y(1-t)](\beta_z t)^2
\]
\[
+ 4z^4[u(1-x)][y(1-t)](\beta_z t)^2
+ 4z^4[u(1-x)][y(1-t)](\beta_z t) - 2z^4[y(1-t)](\beta_z t)^2
\]
\[
- 4z^4[y(1-t)](\beta_z t) - 2z^4[y(1-t)]
+ 2z^3(1-z)(ux)^2[y(1-t)]
\]
\[
+ 4z^3(1-z)(ux)[u(1-x)][y(1-t)](\beta_z t) - 4z^3(1-z)(ux)[y(1-t)](\beta_z t)
\]
\[
- 4z^3(1-z)(ux)[y(1-t)]
+ 2z^3(1-z)[u(1-x)]^2[y(1-t)](\beta_z t)^2
\]
\[
- 4z^3(1-z)[u(1-x)][y(1-t)](\beta_z t)^2
- 4z^3(1-z)[u(1-x)][y(1-t)](\beta_z t)
\]
\[
+ 2z^3(1-z)[y(1-t)](\beta_z t)^2 + 4z^3(1-z)[y(1-t)](\beta_z t)
+ 2z^3(1-z)[y(1-t)]
\]
\[
+ 8z^3(ux)[u(1-x)][y(1-t)](\beta_z t) + 2z^2[y(1-t)](\beta_z t)^2
+ 3z^2[y(1-t)](\beta_z t)
\]
\[
+ z^2[y(1-t)] - z^2(\beta_z t)^2
- z^2(\beta_z t) - z^2 + 2z(ux)^2[y(1-t)]
\]
\[
+ 4z(ux)[u(1-x)][y(1-t)](\beta_z t) - 4z(ux)[y(1-t)](\beta_z t)
- 4z(ux)[y(1-t)]
\]
\[
+ 2z[u(1-x)]^2[y(1-t)](\beta_z t)^2 - 4z[u(1-x)][y(1-t)](\beta_z t)^2
\]
\[
- 4z[u(1-x)][y(1-t)](\beta_z t) + 2z[y(1-t)](\beta_z t)^2
+ 4z[y(1-t)](\beta_z t) + 2z[y(1-t)]
\]
\[
- 2(1-z)(ux)^2[y(1-t)] - 4(1-z)(ux)[u(1-x)][y(1-t)](\beta_z t)
\]
\[
+ 4(1-z)(ux)[y(1-t)](\beta_z t) + 4(1-z)(ux)[y(1-t)]
\]
\[
- 2(1-z)[u(1-x)]^2[y(1-t)](\beta_z t)^2
+ 4(1-z)[u(1-x)][y(1-t)](\beta_z t)^2
\]
\[
+ 4(1-z)[u(1-x)][y(1-t)](\beta_z t) - 2(1-z)[y(1-t)](\beta_z t)^2
- 4(1-z)[y(1-t)](\beta_z t)
\]
\[
- 2(1-z)[y(1-t)] + 4(ux)[u(1-x)][y(1-t)](\beta_z t),
\]

\medskip

\beq
P^c_3=6b^2z^2[y(1-t)] - 6b^2z^2 - 6bz^2[y(1-t)] + 6bz^2
\eeq
\[
- 6s^2z^3(1-z)[y(1-t)] - 4sz^3[u(1-x)][y(1-t)]
- 2s[u(1-x)][y(1-t)]
\]
\[
- 2z^4[u(1-x)]^2[y(1-t)] + 4z^4[u(1-x)][y(1-t)] - 2z^4[y(1-t)]
\]
\[
+ 2z^3(1-z)[u(1-x)]^2[y(1-t)] - 4z^3(1-z)[u(1-x)][y(1-t)] + 2z^3(1-z)[y(1-t)]
\]
\[
- 3z^2 + 2z[u(1-x)]^2[y(1-t)] - 4z[u(1-x)][y(1-t)] + 2z[y(1-t)]
\]
\[
- 2(1-z)[u(1-x)]^2[y(1-t)] + 4(1-z)[u(1-x)][y(1-t)] - 2(1-z)[y(1-t)],
\]

Second term in the integrand in \eq{int2z} has the form

\beq
2F^c_2=2q''^4mT^c_4 - q''^2k^2mT^c_1  - q''^2m^3T^c_5
\eeq
\[
+ 2k^4mT^c_2+ 2k^2m^3T^c_3  + 2m^5[y(1-t)]^2T^c_6,
\]

where \footnote{It is not difficult to check that despite its cumbersome
appearance function $T^c_4$ vanishes identically: $T^c_4\equiv 0$.}

\beq
T^c_4 =  - 4b^4[y(1-t)] + b^4 + 8b^3[y(1-t)] - 2b^3
\eeq
\[
- 4b^2[y(1-t)] + b^2 + 4s^4z^2(1-z)^2[y(1-t)] - s^4z^2(1-z)^2
\]
\[
+ 8s^3z^2(1-z)[u(1-x)][y(1-t)] - 2s^3z^2(1-z)[u(1-x)]
\]
\[
+ 4s^2z^2[u(1-x)]^2[y(1-t)] - s^2z^2[u(1-x)]^2,
\]

\medskip

\beq
T^c_1 =  - 6a^2b^2[y(1-t)] + 6a^2b^2 + 6a^2b[y(1-t)] - 6a^2b
\eeq
\[
- 3a^2[y(1-t)] + 3a^2 - 16ab^3[y(1-t)](\beta_z t)
+ 12ab^3(\beta_z t) + 24ab^2[y(1-t)](\beta_z t)
\]
\[
+ 6ab^2[y(1-t)] - 18ab^2(\beta_z t) - 6ab^2 - 8ab[y(1-t)](\beta_z t)
- 6ab[y(1-t)]
\]
\[
+ 6ab(\beta_z t) + 6ab + 3a[y(1-t)] - 3a - 16b^4[y(1-t)](\beta_z t)^2
+ 6b^4(\beta_z t)^2
\]
\[
+ 32b^3[y(1-t)](\beta_z t)^2
+ 8b^3[y(1-t)](\beta_z t) - 12b^3(\beta_z t)^2 - 6b^3(\beta_z t)
\]
\[
- 16b^2[y(1-t)](\beta_z t)^2 - 12b^2[y(1-t)](\beta_z t)
- 3b^2[y(1-t)] + 6b^2(\beta_z t)^2 + 9b^2(\beta_z t)
\]
\[
+ 2b^2 + 4b[y(1-t)](\beta_z t) + 3b[y(1-t)]
- 3b(\beta_z t) - 2b + 16s^4z^2(1-z)^2[y(1-t)](\beta_z t)^2
\]
\[
- 6s^4z^2(1-z)^2(\beta_z t)^2 + 32s^3vz^2(1-z)^2[y(1-t)](\beta_z t)
- 12s^3vz^2(1-z)^2(\beta_z t)
\]
\[
+ 32s^3z^2(1-z)[u(1-x)][y(1-t)](\beta_z t)^2
- 12s^3z^2(1-z)[u(1-x)](\beta_z t)^2
\]
\[
+ 16s^2v^2z^2(1-z)^2[y(1-t)] - 6s^2v^2z^2(1-z)^2
+ 12s^2vz^2(1-z)(ux)[y(1-t)]
\]
\[
- 4s^2vz^2(1-z)(ux) + 32s^2vz^2(1-z)[u(1-x)][y(1-t)](\beta_z t)
\]
\[
- 12s^2vz^2(1-z)[u(1-x)](\beta_z t)
- 32s^2z^2(1-z)(ux)[u(1-x)][y(1-t)](\beta_z t)
\]
\[
+ 12s^2z^2(1-z)(ux)[u(1-x)](\beta_z t)
+ 16s^2z^2[u(1-x)]^2[y(1-t)](\beta_z t)^2
\]
\[
- 6s^2z^2[u(1-x)]^2(\beta_z t)^2 + 12sv^2z^2(1-z)[u(1-x)][y(1-t)]
\]
\[
- 4sv^2z^2(1-z)[u(1-x)] - 8svz^2(1-z)(ux)[u(1-x)][y(1-t)]
\]
\[
+ 4svz^2(1-z)(ux)[u(1-x)] + 12svz^2(ux)[u(1-x)][y(1-t)]
\]
\[
- 4svz^2(ux)[u(1-x)]- 32sz^2(ux)[u(1-x)]^2[y(1-t)](\beta_z t)
\]
\[
+ 12sz^2(ux)[u(1-x)]^2(\beta_z t) + 4z^2(ux)^2[u(1-x)]^2[y(1-t)]
\]
\[
- 2z^2(ux)^2[u(1-x)]^2,
\]

\medskip

\beq
T^c_5 =  - 16b^4[y(1-t)]^3 + 6b^4[y(1-t)]^2 + 32b^3[y(1-t)]^3
\eeq
\[
- 12b^3[y(1-t)]^2 - 16b^2[y(1-t)]^3
+ 6b^2[y(1-t)]^2 + 12b^2[y(1-t)] - 4b^2
\]
\[
- 12b[y(1-t)] + 4b + 16s^4z^2(1-z)^2[y(1-t)]^3 - 6s^4z^2(1-z)^2[y(1-t)]^2
\]
\[
+ 32s^3z^2(1-z)[u(1-x)][y(1-t)]^3- 12s^3z^2(1-z)[u(1-x)][y(1-t)]^2
\]
\[
- 6s^2z^2(1-z)[y(1-t)] + 2s^2z^2(1-z) + 16s^2z^2[u(1-x)]^2[y(1-t)]^3
\]
\[
- 6s^2z^2[u(1-x)]^2[y(1-t)]^2 - 6s^2z(1-z)^2[y(1-t)] + 2s^2z(1-z)^2
\]
\[
- 6s^2z(1-z)[y(1-t)] + 2s^2z(1-z)
- 6sz^2[u(1-x)][y(1-t)] + 2sz^2[u(1-x)]
\]
\[
- 6sz(1-z)[u(1-x)][y(1-t)] + 2sz(1-z)[u(1-x)] - 6sz[u(1-x)][y(1-t)]
\]
\[
+ 2sz[u(1-x)] + 3[y(1-t)] - 3,
\]

\medskip

\beq
T^c_2= - 2a^4[y(1-t)] + a^4 - 4a^3b[y(1-t)](\beta_z t) + 4a^3b(\beta_z t)
\eeq
\[
+ 2a^3[y(1-t)](\beta_z t) + 4a^3[y(1-t)]
- 2a^3(\beta_z t) - 2a^3 - 4a^2b^2[y(1-t)](\beta_z t)^2
\]
\[
+ 6a^2b^2(\beta_z t)^2 + 4a^2b[y(1-t)](\beta_z t)^2
+ 6a^2b[y(1-t)](\beta_z t) - 6a^2b(\beta_z t)^2 - 6a^2b(\beta_z t)
\]
\[
- 2a^2[y(1-t)](\beta_z t)^2 - 3a^2[y(1-t)](\beta_z t)
- 2a^2[y(1-t)] + a^2(\beta_z t)^2 + 3a^2(\beta_z t)
\]
\[
+ a^2 - 4ab^3[y(1-t)](\beta_z t)^3
+ 4ab^3(\beta_z t)^3 + 6ab^2[y(1-t)](\beta_z t)^3
+ 4ab^2[y(1-t)](\beta_z t)^2
\]
\[
- 6ab^2(\beta_z t)^3 - 6ab^2(\beta_z t)^2 - 2ab[y(1-t)](\beta_z t)^3
- 4ab[y(1-t)](\beta_z t)^2
\]
\[
- 2ab[y(1-t)](\beta_z t) + 2ab(\beta_z t)^3 + 6ab(\beta_z t)^2
+ 2ab(\beta_z t) + 2a[y(1-t)](\beta_z t)^2
\]
\[
+ a[y(1-t)](\beta_z t) - a(\beta_z t)^2 - a(\beta_z t)
- 2b^4[y(1-t)](\beta_z t)^4 + b^4(\beta_z t)^4
\]
\[
+ 4b^3[y(1-t)](\beta_z t)^4 + 2b^3[y(1-t)](\beta_z t)^3 - 2b^3(\beta_z t)^4
- 2b^3(\beta_z t)^3
\]
\[
- 2b^2[y(1-t)](\beta_z t)^4 - 3b^2[y(1-t)](\beta_z t)^3
- b^2[y(1-t)](\beta_z t)^2 + b^2(\beta_z t)^4 + 3b^2(\beta_z t)^3
\]
\[
+ b^2(\beta_z t)^2 + b[y(1-t)](\beta_z t)^3 + b[y(1-t)](\beta_z t)^2
- b(\beta_z t)^3 - b(\beta_z t)^2\]
\[
+ 2s^4z^2(1-z)^2[y(1-t)](\beta_z t)^4
- s^4z^2(1-z)^2(\beta_z t)^4 + 8s^3vz^2(1-z)^2[y(1-t)](\beta_z t)^3
\]
\[
- 4s^3vz^2(1-z)^2(\beta_z t)^3 + 4s^3z^2(1-z)[u(1-x)][y(1-t)](\beta_z t)^4
\]
\[
- 2s^3z^2(1-z)[u(1-x)](\beta_z t)^4 + 12s^2v^2z^2(1-z)^2[y(1-t)](\beta_z t)^2
\]
\[
- 6s^2v^2z^2(1-z)^2(\beta_z t)^2 + 4s^2vz^2(1-z)(ux)[y(1-t)](\beta_z t)^2
\]
\[
- 2s^2vz^2(1-z)(ux)(\beta_z t)^2 + 8s^2vz^2(1-z)[u(1-x)][y(1-t)](\beta_z t)^3
\]
\[
- 4s^2vz^2(1-z)[u(1-x)](\beta_z t)^3
- 8s^2z^2(1-z)(ux)[u(1-x)][y(1-t)](\beta_z t)^3
\]
\[
+ 4s^2z^2(1-z)(ux)[u(1-x)](\beta_z t)^3
+ 2s^2z^2[u(1-x)]^2[y(1-t)](\beta_z t)^4
\]
\[
- s^2z^2[u(1-x)]^2(\beta_z t)^4 + 8sv^3z^2(1-z)^2[y(1-t)](\beta_z t)
- 4sv^3z^2(1-z)^2(\beta_z t)
\]
\[
+ 8sv^2z^2(1-z)(ux)[y(1-t)](\beta_z t)
- 4sv^2z^2(1-z)(ux)(\beta_z t)
\]
\[
+ 4sv^2z^2(1-z)[u(1-x)][y(1-t)](\beta_z t)^2
- 2sv^2z^2(1-z)[u(1-x)](\beta_z t)^2
\]
\[
- 16svz^2(1-z)(ux)[u(1-x)][y(1-t)](\beta_z t)^2
+ 8svz^2(1-z)(ux)[u(1-x)](\beta_z t)^2
\]
\[
+ 4svz^2(ux)[u(1-x)][y(1-t)](\beta_z t)^2
- 2svz^2(ux)[u(1-x)](\beta_z t)^2
\]
\[
- 8sz^2(ux)[u(1-x)]^2[y(1-t)](\beta_z t)^3
+ 4sz^2(ux)[u(1-x)]^2(\beta_z t)^3
\]
\[
+ 2v^4z^2(1-z)^2[y(1-t)]
- v^4z^2(1-z)^2 + 4v^3z^2(1-z)(ux)[y(1-t)]
\]
\[
- 2v^3z^2(1-z)(ux) - 8v^2z^2(1-z)(ux)[u(1-x)][y(1-t)](\beta_z t)
\]
\[
+ 4v^2z^2(1-z)(ux)[u(1-x)](\beta_z t) + 2v^2z^2(ux)^2[y(1-t)] - v^2z^2(ux)^2
\]
\[
- 8vz^2(ux)^2[u(1-x)][y(1-t)](\beta_z t)
+ 4vz^2(ux)^2[u(1-x)](\beta_z t)
\]
\[
+ 8z^2(ux)^2[u(1-x)]^2[y(1-t)](\beta_z t)^2
- 4z^2(ux)^2[u(1-x)]^2(\beta_z t)^2,
\]

\medskip

\beq
T^c_3=2a^2b^2[y(1-t)]^2 - 2a^2b[y(1-t)]^2 + 3a^2[y(1-t)]^2
\eeq
\[
+ 4a^2[y(1-t)] - 2a^2
- 4ab^3[y(1-t)]^3(\beta_z t) + 4ab^3[y(1-t)]^2(\beta_z t)
\]
\[
+ 6ab^2[y(1-t)]^3(\beta_z t) - 6ab^2[y(1-t)]^2(\beta_z t)
- 2ab^2[y(1-t)]^2 - 2ab[y(1-t)]^3(\beta_z t)
\]
\[
+ 2ab[y(1-t)]^2(\beta_z t) + 2ab[y(1-t)]^2
+ 4ab[y(1-t)](\beta_z t) - 4ab(\beta_z t) - 3a[y(1-t)]^2
\]
\[
- 2a[y(1-t)](\beta_z t) - 4a[y(1-t)] + 2a(\beta_z t) + 2a
- 4b^4[y(1-t)]^3(\beta_z t)^2
\]
\[
+ 2b^4[y(1-t)]^2(\beta_z t)^2 + 8b^3[y(1-t)]^3(\beta_z t)^2
+ 2b^3[y(1-t)]^3(\beta_z t)
\]
\[
- 4b^3[y(1-t)]^2(\beta_z t)^2 - 2b^3[y(1-t)]^2(\beta_z t)
- 4b^2[y(1-t)]^3(\beta_z t)^2
\]
\[
- 3b^2[y(1-t)]^3(\beta_z t)
- b^2[y(1-t)]^3 + 2b^2[y(1-t)]^2(\beta_z t)^2 + 3b^2[y(1-t)]^2(\beta_z t)
\]
\[
+ b^2[y(1-t)]^2 + 4b^2[y(1-t)](\beta_z t)^2 - 2b^2(\beta_z t)^2
+ b[y(1-t)]^3(\beta_z t) + b[y(1-t)]^3
\]
\[
- b[y(1-t)]^2(\beta_z t) - b[y(1-t)]^2 - 4b[y(1-t)](\beta_z t)^2
- 2b[y(1-t)](\beta_z t) + 2b(\beta_z t)^2
\]
\[
+ 2b(\beta_z t) + 4s^4z^2(1-z)^2[y(1-t)]^3(\beta_z t)^2
- 2s^4z^2(1-z)^2[y(1-t)]^2(\beta_z t)^2
\]
\[
+ 8s^3vz^2(1-z)^2[y(1-t)]^3(\beta_z t)
- 4s^3vz^2(1-z)^2[y(1-t)]^2(\beta_z t)
\]
\[
+ 8s^3z^2(1-z)[u(1-x)][y(1-t)]^3(\beta_z t)^2
- 4s^3z^2(1-z)[u(1-x)][y(1-t)]^2(\beta_z t)^2
\]
\[
+ 4s^2v^2z^2(1-z)^2[y(1-t)]^3
- 2s^2v^2z^2(1-z)^2[y(1-t)]^2
\]
\[
+ 4s^2vz^2(1-z)(ux)[y(1-t)]^3 - 2s^2vz^2(1-z)(ux)[y(1-t)]^2
\]
\[
+ 8s^2vz^2(1-z)[u(1-x)][y(1-t)]^3(\beta_z t)
- 4s^2vz^2(1-z)[u(1-x)][y(1-t)]^2(\beta_z t)
\]
\[
- 8s^2z^2(1-z)(ux)[u(1-x)][y(1-t)]^3(\beta_z t)
\]
\[
+ 4s^2z^2(1-z)(ux)[u(1-x)][y(1-t)]^2(\beta_z t)
- 2s^2z^2(1-z)[y(1-t)](\beta_z t)^2
\]
\[
+ s^2z^2(1-z)(\beta_z t)^2 + 4s^2z^2[u(1-x)]^2[y(1-t)]^3(\beta_z t)^2
\]
\[
- 2s^2z^2[u(1-x)]^2[y(1-t)]^2(\beta_z t)^2
- 2s^2z(1-z)^2[y(1-t)](\beta_z t)^2 + s^2z(1-z)^2(\beta_z t)^2
\]
\[
- 2s^2z(1-z)[y(1-t)](\beta_z t)^2
+ s^2z(1-z)(\beta_z t)^2 + 4sv^2z^2(1-z)[u(1-x)][y(1-t)]^3
\]
\[
- 2sv^2z^2(1-z)[u(1-x)][y(1-t)]^2
- 4svz^2(1-z)[y(1-t)](\beta_z t) + 2svz^2(1-z)(\beta_z t)
\]
\[
+ 4svz^2(ux)[u(1-x)][y(1-t)]^3
- 2svz^2(ux)[u(1-x)][y(1-t)]^2
\]
\[
- 4svz(1-z)^2[y(1-t)](\beta_z t) + 2svz(1-z)^2(\beta_z t)
- 4svz(1-z)[y(1-t)](\beta_z t)
\]
\[
+ 2svz(1-z)(\beta_z t) - 8sz^2(ux)[u(1-x)]^2[y(1-t)]^3(\beta_z t)
\]
\[
+ 4sz^2(ux)[u(1-x)]^2[y(1-t)]^2(\beta_z t)
- 2sz^2[u(1-x)][y(1-t)](\beta_z t)^2
\]
\[
+ sz^2[u(1-x)](\beta_z t)^2 - 2sz(1-z)[u(1-x)][y(1-t)](\beta_z t)^2
\]
\[
+ sz(1-z)[u(1-x)](\beta_z t)^2 - 2sz[u(1-x)][y(1-t)](\beta_z t)^2
\]
\[
+ sz[u(1-x)](\beta_z t)^2 - 2v^2z^2(1-z)[y(1-t)] + v^2z^2(1-z)
\]
\[
- 2v^2z(1-z)^2[y(1-t)] + v^2z(1-z)^2 - 2v^2z(1-z)[y(1-t)] + v^2z(1-z)
\]
\[
- 2vz^2(ux)[y(1-t)] + vz^2(ux) - 2vz(1-z)(ux)[y(1-t)] + vz(1-z)(ux)
\]
\[
- 2vz(ux)[y(1-t)] + vz(ux) + 4z^2(ux)[u(1-x)][y(1-t)](\beta_z t)
\]
\[
- 2z^2(ux)[u(1-x)](\beta_z t) + 4z(1-z)(ux)[u(1-x)][y(1-t)](\beta_z t)
\]
\[
- 2z(1-z)(ux)[u(1-x)](\beta_z t) + 4z(ux)[u(1-x)][y(1-t)](\beta_z t)
\]
\[
- 2z(ux)[u(1-x)](\beta_z t) + 2[y(1-t)](\beta_z t)^2
+ [y(1-t)](\beta_z t) + [y(1-t)]
\]
\[
- (\beta_z t)^2 - (\beta_z t) - 1,
\]

\medskip

\medskip

\beq
T^c_6 = - 2b^4[y(1-t)]^3 + b^4[y(1-t)]^2 + 4b^3[y(1-t)]^3
\eeq
\[
- 2b^3[y(1-t)]^2 - 2b^2[y(1-t)]^3
+ b^2[y(1-t)]^2 + 4b^2[y(1-t)] - 2b^2 - 4b[y(1-t)]
\]
\[
+ 2b + 2s^4z^2(1-z)^2[y(1-t)]^3 - s^4z^2(1-z)^2[y(1-t)]^2
\]
\[
+ 4s^3z^2(1-z)[u(1-x)][y(1-t)]^3 - 2s^3z^2(1-z)[u(1-x)][y(1-t)]^2
\]
\[
- 2s^2z^2(1-z)[y(1-t)] + s^2z^2(1-z) + 2s^2z^2[u(1-x)]^2[y(1-t)]^3
\]
\[
- s^2z^2[u(1-x)]^2[y(1-t)]^2 - 2s^2z(1-z)^2[y(1-t)] + s^2z(1-z)^2
\]
\[
- 2s^2z(1-z)[y(1-t)] + s^2z(1-z)
- 2sz^2[u(1-x)][y(1-t)] + sz^2[u(1-x)]
\]
\[
- 2sz(1-z)[u(1-x)][y(1-t)] + sz(1-z)[u(1-x)]
- 2sz[u(1-x)][y(1-t)]
\]
\[
+ sz[u(1-x)]  - 3.
\]

After integration we obtain

\beq            \label{infraredc}
\Delta E_{c}=\frac{\alpha^2(Z\alpha)^5}{\pi n^3}m
(\frac{m_r}{m})^3\frac{12}{\pi^2}
\int_0^1dx\int_0^1duu\int_0^1dz\gamma_z
\eeq
\[
\int_0^1dy(1-y)\int_0^1dt(1-t)^2\int\frac{d|\bf k|}{{\bf k}^2}
\{4\gamma_z t{\bf k}^4\frac{mz^2T^c_2}{6d_{\lambda z}^6}
\]
\[
-{\bf k}^2[\frac{mP^c_2-\gamma_z tmz^2T^c_1}{3d_{\lambda z}^4}+4\gamma_z t
\frac{m^3z^2T^c_3}{6d_{\lambda z}^6}]
\]
\[
-\frac{3mP^c_1}{3d_{\lambda z}^2}+\frac{m^3[y(1-t)]^2P^c_3}{3d_{\lambda z}^4}
+ 4\gamma_z t\frac{mz^2T^c_4}{2d_{\lambda z}^2}
\]
\[
-2\gamma_z t\frac{m^3z^2T^c_5}{6d_{\lambda z}^4}
+4\gamma_z t\frac{m^5z^2[y(1-t)]^2T^c_6}{6d_{\lambda z}^6} \}.
\]

Integration over $|{\bf k}|$ is no more difficult than in the ladder case.
We introduce notation which resembles the one used in the ladder case

\beq
d_{\lambda z}^2={\bf k}^2t(\alpha_z-\beta_z^2t)+m^2[y^2(1-t)^2+\gamma_z t]
\eeq
\[
\equiv\rho_z \{{\bf k}^2+\frac{m^2[y^2(1-t)^2+\gamma_z
t]+\lambda^2(1-t)(1-y)}{\rho_z}\} \] \[ =\rho_z ({\bf k}^2+\omega_{\lambda
z}^2), \]

where

\beq
\rho_z\equiv t(\alpha_z-\beta_z^2t),
\eeq
\[
\omega_{\lambda z}^2\equiv \frac{m^2[y^2(1-t)^2+\gamma_z
t]+\lambda^2(1-t)(1-y)}{\rho_z}.
\]

Once again as in the ladder case we encounter the problem of apparent
divergence of the integral for the energy shift. As was already explained it
is necessary to check that the total electron-line factor vanishes as ${\bf
k}^2$ in the low frequency region. It is an easy task to do. We  simply omit
integration over ${\bf k}$ in \eq{infrared} and \eq{infraredc} and put ${\bf
k}$ to be equal to zero in the integrand. Then we obtain

\beq
\Delta\epsilon_{test}=\Delta\epsilon _{l}+\Delta
\epsilon_{c}
\eeq
\[
=\int_0^1dx\int_0^1du\int_0^1dy(1-y)\int_0^1dt(1-t)^2
\]
\[
\{\gamma\{-\frac{(1-u)P^l_1}{3[y^2(1-t)^2+\gamma t]}
+\frac{[y(1-t)]^2(1-u)P^l_3}{3[y^2(1-t)^2+\gamma t]^2}
+ 4\gamma t\frac{[u(1-x)]^2u(1-u)T^l_4}{6[y^2(1-t)^2+\gamma t]}
\]
\[
-4\gamma t\frac{[u(1-x)]u(1-u)T^l_5}{6[y^2(1-t)^2+\gamma t]^2}
+4\gamma t\frac{[u(1-x)]u[y(1-t)]^2(1-u)T^l_6}{3[y^2(1-t)^2+\gamma t]^3}\}
\]
\[
+\int_0^1dz\gamma_z
\{-\frac{3uP^c_1}{3[y^2(1-t)^2+\gamma_z t]}
+\frac{u[y(1-t)]^2P^c_3}{3[y^2(1-t)^2+\gamma_z t]^2}
\]
\[
+ 4\gamma_z t\frac{z^2uT^c_4}{2[y^2(1-t)^2+\gamma_z t]}
-2\gamma_z t\frac{z^2uT^c_5}{6[y^2(1-t)^2+\gamma_z t]^2}
\]
\[
+4\gamma_z t\frac{z^2u[y(1-t)]^2T^c_6}{6[y^2(1-t)^2+\gamma_z t]^3}\} \}.
\]

We calculated this integral numerically and it turned out to be equal zero
as expected.  Then there are really no difficulties in the treatment of the
integrals
contained in \eq{infraredc}; one has simply to perform subtraction of the
leading low frequency terms in the electron-line factor since these terms
cancel in any case in the total electron-line factor.

Subtraction of the low frequency part is necessary in the last five terms
of the integrand in \eq{infraredc}. The numerators of these terms are
independent of momentum $k$ and subtraction may be easily performed
with the help of the identities

\beq
\frac{1}{d_{\lambda z}^2({\bf k})}-\frac{1}{d_{\lambda z}^2(0)}
=-\frac{{\bf k}^2}{\omega_{\lambda z}^2d_{\lambda z}^2({\bf
k})}\equiv -\frac{{\bf k}^2}{\rho_z\omega_{\lambda z}^2
({\bf k}^2+\omega_{\lambda z}^2)},
\eeq
\[
\frac{1}{d_{\lambda z}^4({\bf k})}-\frac{1}{d_{\lambda z}^4(0)}=-\frac{{\bf
k}^2}{\rho_z\omega_{\lambda z}^4d_{\lambda z}^2({\bf k})}-\frac{{\bf
k}^2}{\omega_{\lambda z}^2d_z^4({\bf k})}
\]
\[
\equiv -\frac{{\bf k}^2}{\rho_z^2\omega_{\lambda z}^4({\bf
k}^2+\omega_{\lambda z}^2)}-\frac{{\bf k}^2}{\rho_z^2\omega_{\lambda z}^2
({\bf k}^2+\omega_{\lambda z}^2)^2},
\]
\[
\frac{1}{d_{\lambda z}^6({\bf k})}-\frac{1}{d_{\lambda z}^6(0)}=-\frac{{\bf
k}^2}{\rho_z^2\omega_{\lambda z}^6d_{\lambda z}^2({\bf k})}-\frac{{\bf
k}^2}{\rho_z\omega_{\lambda z}^4d_{\lambda z}^4({\bf k})}-\frac{{\bf
k}^2}{\omega_z^2d_{\lambda z}^6({\bf k})}
\]
\[
\equiv -\frac{{\bf k}^2}{\rho_z^3\omega_{\lambda z}^6({\bf
k}^2+\omega_{\lambda z}^2)}-\frac{{\bf k}^2}{\rho_z^3\omega_{\lambda z}^4
({\bf k}^2+\omega_{\lambda z}^2)^2}-\frac{{\bf k}^2}{\rho_z^3
\omega_{\lambda z}^2({\bf k}^2+\omega_{\lambda z}^2)^3}.
\]

Substituting these subtractions in  \eq{infraredc} and performing the momentum
integration we obtain

\beq       \label{crossedcontr}
\Delta E_{c}^{sub}=\frac{\alpha^2(Z\alpha)^5}{\pi n^3}m
(\frac{m_r}{m})^3\frac{12}{\pi}
\int_0^1dx
\eeq
\[
\int_0^1duu\int_0^1dz\gamma_z\int_0^1dy(1-y)\int_0^1dt(1-t)^2
\]
\[
\{\frac{\gamma_z t}{8\rho_z^3\omega_{\lambda z}^5}
[-m\omega_{\lambda z}^2z^2T^c_2
-m^3z^2T^c_3-\frac{m^5z^2[y(1-t)]^2T^c_6}{\omega_{\lambda z}^2}]
\]
\[
+\frac{1}{12\rho_z^2\omega_{\lambda z}^3}[
2\gamma_z t\frac{mz^2T^c_2}{\rho_z}
-mP^c_2+\gamma_z tmz^2T^c_1
-\frac{m^3[y(1-t)]^2P^c_3}{\omega_{\lambda z}^2}
\]
\[
+\gamma_z t\frac{m^3z^2T^c_5}{\omega_{\lambda z}^2}
-2\gamma_z t\frac{m^5z^2[y(1-t)]^2T^c_6}{\rho_z\omega_{\lambda z}^4}]
\]
\[
+\frac{m}{6\rho_z\omega_{\lambda z}^3}
[3P^c_1
-\frac{m^2[y(1-t)]^2P^c_3}{\rho_z\omega_{\lambda z}^2}
- 6\gamma_z tz^2T^c_4
\]
\[
+\gamma_z t\frac{m^2z^2T^c_5}{\rho_z\omega_{\lambda z}^2}
-2\gamma_z
t\frac{m^4z^2[y(1-t)]^2T^c_6}{\rho_z^2\omega_{\lambda z}^4}] \}.
\]

\section{TOTAL CONTRIBUTION TO THE LAMB SHIFT AND DISCUSSION OF RESULTS}

It is straightforward to calculate numerically the total contribution to
the $S$-level Lamb shift of order $\alpha^2(Z\alpha)^5m$ induced by the
light by light scattering insertion. This contribution is given by the
sum of the expressions in \eq{laddercontr}  and in \eq{crossedcontr}. The
only subtlety is that these expressions contain an auxiliary infrared
regularizing parameter $\lambda$. As we have mentioned above  the final formula
for the energy shift should be given in the limit of the vanishing photon
mass and should
produce an unambiguous value for the contribution to the Lamb shift. We have
checked numerically that the result of integration is independent
of this small intermediate photon mass as it is varied from $\lambda^2=10^{-4}$
to $\lambda^2=10^{-8}$. This limited range of values for the
auxiliary photon mass is defined by the limited accuracy and productivity
of the computer and may be widened at the expense of significantly more
computer
time. Our final result is

\beq
\Delta E=-0.122(2)\;\frac{\alpha^2(Z\alpha)^5}{\pi
n^3}\;(\frac{m_r}{m})^3\;m.
\eeq

One may readily obtain more digits in the expression above if necessary.

In this work, initiated in \cite{ego}, we have finished calculation of all
corrections to
the Lamb shift of order $\alpha^2(Z\alpha)^5m$ induced by
the diagrams with closed electron loops.
Our results are presented in the Table.\footnote {While this paper
was in preparation a work \cite{pach} appeared which  contains
recalculation and confirmation of the results obtained in \cite{ego,eg3,eg4}
as well as the first calculation of the light by light contribution in a
framework completely different from the one used in this
paper and in \cite{ego,eg3,eg4}. The result for the light by light
contribution presented above is in agreement with the respective
result in \cite{pach}.}.

\begin{center}
Corrections to the Lamb shift\nopagebreak\\ induced by the diagrams
with closed electron loops
\nopagebreak \vspace*{4pt}

\begin{tabular}{|l|rl|r|r|}    \hline

\ $\Delta E$  & $\times\frac{\alpha^2(Z\alpha)^5}{\pi
n^3}(\frac{m_r}{m})^3m$ & &$2S$ (kHz) &$1S$  (kHz)
\\ \hline
$a^{\cite{ego}}$&   $- 0.061$    &        &  $-0.329$ &  $-2.63$
\\ \hline
$b^{\cite{ego}}$&   $0.508$      &        & $2.747$  &  $21.98$
\\ \hline
$c^{\cite{eg3}}$    &   $0.611$       &           & $3.305$ &  $26.44$
\\ \hline
$d^{\cite{eg4}}$    &   $-0.073$      &        & $-0.394$ &  $-3.15$
\\ \hline
e-\sl{this work}  &   $-0.122(2)$       &        & $-0.660$  &  $-5.28$
\\ \hline
Total  &   $0.863(2)$       &        & $4.67(1)$  &  $37.3(1)$\\ \hline
\end{tabular}
\end{center}

\hspace{2pt}

The theoretical contributions to the Lamb shift presented in the Table are
clearly necessary for comparison of theory with recent experiments
on measurement of the $2S_{1/2} - 2P_{1/2}$ splitting in hydrogen
\cite{lp,sok}

\[
\Delta \nu  = 1057\; 845\; (9)\hbox{ kHz}\:,
\]
\beq
\Delta \nu  = 1057\;851.4\; (1.9)\hbox{ kHz},
\eeq
and on measurement of the $1S$ Lamb shift
\cite{weitz}

\beq
\Delta \nu_{1S} = 8\; 172.82\;(11) \hbox{ MHz}.
\eeq

Work on the calculation of the still unknown corrections of order
$\alpha^2(Z\alpha)^5m$ produced by the diagrams with two radiative
photon insertions in the electron line is in the
finishing stage now by at least two groups\footnote{K. Pachucki, submitted
for publication; M. I.  Eides, S.  G. Karshenboim and V. A. Shelyuto
\cite{ekse1,ekse2} and paper in preparation.}.

\bigskip
We are deeply grateful to Dr. K. Pachucki for informing us about his results
prior to publication. It is a pleasure for us to acknowledge the help of
Dr. Shoudan Liang with numerical calculations. This work was partially
fulfilled during the visit by one of the authors (M.E.) to the Penn State
University. He is deeply grateful to the Physics Department of the Penn
State University for their kind hospitality.

\bigskip
This research was supported by the National Science Foundation
under grant \#NSF-PHY-9120102. P.P received support from NSF through
funds provided by the site REU program at Penn State.  Work of M. E. was also
partially supported by
the Russian Foundation for Fundamental Research under grant \#93-02-3853.

\newpage
\begin{center}\large Figure Caption
\end{center}
\vskip5cm
Fig.1. Gauge invariant set of diagrams with light by light scattering
insertions in the external Coulomb lines.

\newpage
\begin{figure}
 \begin{picture}(100,380)
  \put(80,240){\begin{picture}(200,240)
             \thicklines
             \put(-150,170){\line(1,0){140}}

             \multiput(-110,97.5)(0,10){8}{\oval(5,5)[ll]}
             \multiput(-110,92.5)(0,10){8}{\oval(5,5)[rr]}
             \multiput(-50,97.5)(0,10){8}{\oval(5,5)[ll]}
             \multiput(-50,92.5)(0,10){8}{\oval(5,5)[rr]}

             \put(-110,90){\line(1,0){60}}
             \put(-110,90){\line(0,-1){60}}
             \put(-50,90){\line(0,-1){60}}
             \multiput(-110,30)(0,90){1}{\line(1,0){60}}
             \multiput(-110,27.5)(0,-10){8}{\oval(5,5)[ll]}
             \multiput(-110,22.5)(0,-10){8}{\oval(5,5)[rr]}
             \multiput(-50,27.5)(0,-10){8}{\oval(5,5)[ll]}
             \multiput(-50,22.5)(0,-10){8}{\oval(5,5)[rr]}
             \put(-85,-80){\sl{a}}
             \put(0,60){$+$}

             \put(10,170){\line(1,0){140}}

             \multiput(50,97.5)(0,10){8}{\oval(5,5)[ll]}
             \multiput(50,92.5)(0,10){8}{\oval(5,5)[rr]}
             \multiput(110,97.5)(0,10){8}{\oval(5,5)[ll]}
             \multiput(110,92.5)(0,10){8}{\oval(5,5)[rr]}

             \put(50,90){\line(1,0){60}}
             \put(50,90){\line(1,-1){60}}
             \put(110,90){\line(-1,-1){60}}
             \multiput(50,30)(0,90){1}{\line(1,0){60}}
             \multiput(50,27.5)(0,-10){8}{\oval(5,5)[ll]}
             \multiput(50,22.5)(0,-10){8}{\oval(5,5)[rr]}
             \multiput(110,27.5)(0,-10){8}{\oval(5,5)[ll]}
             \multiput(110,22.5)(0,-10){8}{\oval(5,5)[rr]}

             \put(75,-80){\sl{b}}
             \put(150,60){$+$}

             \put(170,170){\line(1,0){140}}

             \multiput(210,97.5)(0,10){8}{\oval(5,5)[ll]}
             \multiput(210,92.5)(0,10){8}{\oval(5,5)[rr]}
             \multiput(270,97.5)(0,10){8}{\oval(5,5)[ll]}
             \multiput(270,92.5)(0,10){8}{\oval(5,5)[rr]}

             \put(210,90){\line(1,-1){60}}
             \put(210,90){\line(0,-1){60}}
             \put(270,90){\line(0,-1){60}}
             \multiput(210,30)(0,90){1}{\line(1,1){60}}
             \multiput(210,27.5)(0,-10){8}{\oval(5,5)[ll]}
             \multiput(210,22.5)(0,-10){8}{\oval(5,5)[rr]}
             \multiput(270,27.5)(0,-10){8}{\oval(5,5)[ll]}
             \multiput(270,22.5)(0,-10){8}{\oval(5,5)[rr]}

             \put(240,-80){\sl{c}}

             \put(300,60){$+$}

             \put(320,70){\sl{Crossed}}
             \put(320,50){\sl{diagrams}}

             \put(110,-120){Fig.1}
             \end{picture}}
\end{picture}
\end{figure}


\newpage \begin{thebibliography}{99}

\bibitem{ego} M. I. Eides, H. Grotch and D. A. Owen, Phys. Lett. B294
   (1992) 115.
\bibitem{eg3} M. I. Eides and H. Grotch, Phys. Lett. B301 (1993) 127.
\bibitem{eg4} M. I. Eides and H. Grotch, Phys. Lett. B308 (1993) 389.
\bibitem{akhb} A. I. Akhiezer and V. B. Berestetsky, Quantum
Electrodynamics, Moscow, Nauka Publishers, 1969 /in Russian/.
\bibitem{pach} K. Pachucki, Phys. Rev. A48 (1993) 2609.
\bibitem{lp} S. R. Lundeen and F. M. Pipkin, Phys. Rev. Lett. 46 (1981) 232;
Metrologia 22 (1986) 9.
\bibitem{sok} Yu. L. Sokolov, and V. P. Yakovlev, Zh. Eksp. Teor. Fiz.
83 (1982) 15 /in Russian/; V. G. Palchikov, Yu. L. Sokolov and V. P.
Yakovlev, Pis'ma Zh. Eksp. Teor. Fiz. 38 (1983) 347 /in Russian/.
\bibitem{weitz} M. Weitz, F. Schmidt-Kaler, and T. W. Hansch, Phys. Rev.
Lett. 68 (1992) 1120.
\bibitem{ekse1} M. I. Eides, S. G. Karshenboim and V. A. Shelyuto,
Phys. Lett. B312 (1993) 358; M. I. Eides, S. G. Karshenboim and V. A.
Shelyuto, Penn State preprint PSU/TH/129, June 1993.
\bibitem{ekse2} M. I. Eides, S. G. Karshenboim and V. A. Shelyuto,
Petersburg Nuclear Physics Institute preprint PNPI-1942/hep-ph-9312337,
December 1993.


\end{thebibliography}
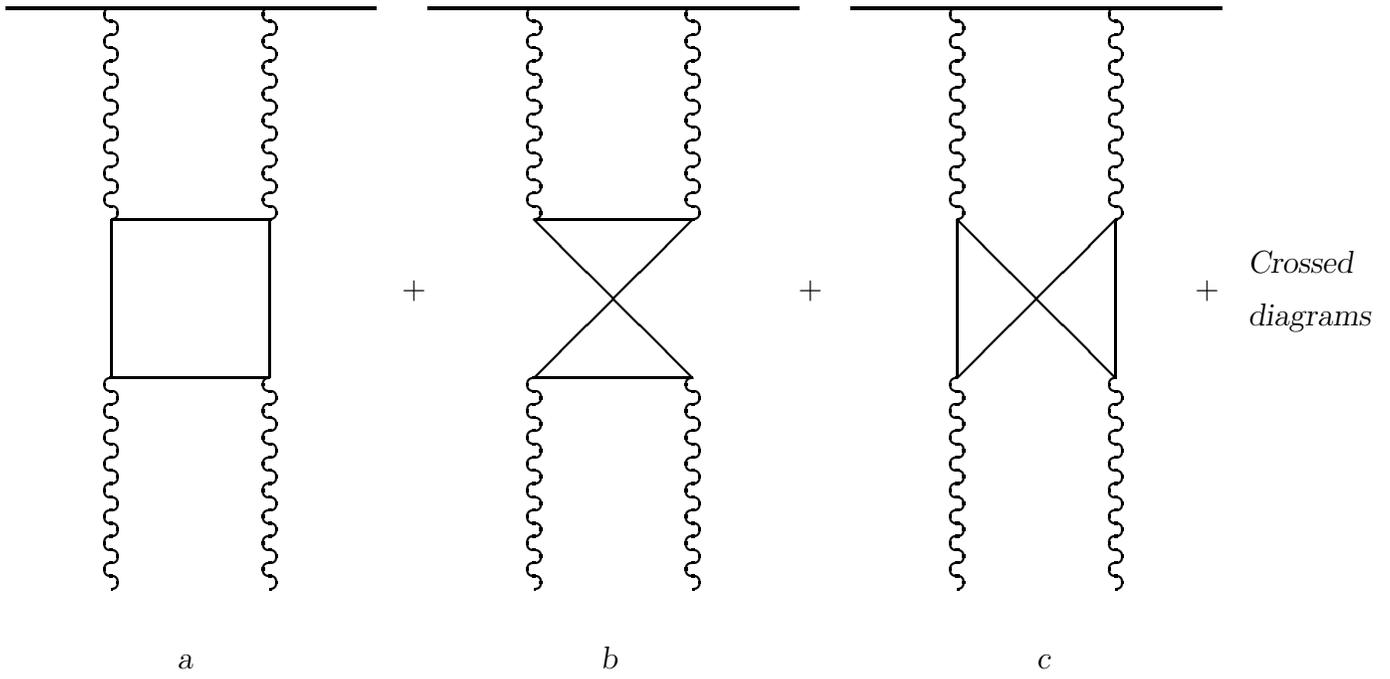
\end{document}